\documentclass[pra,showpacs,floatfix,twocolumn,nofootinbib]{revtex4}

\usepackage{amsmath}
\usepackage{graphicx}
\usepackage{amsfonts}
\usepackage{amssymb}

\newtheorem{mydefinition}{Definition}
\newtheorem{mytheorem}{Theorem}

\newtheorem{mycorollary}{Corollary}

\begin{document}

\title{Theory of Initialization-Free Decoherence-Free Subspaces and
Subsystems}
\author{Alireza Shabani$^{(1)}$ and Daniel A. Lidar$^{(2)}$}
\affiliation{$^{(1)}$Physics Department, and Center for Quantum Information and Quantum
Control, University of Toronto, 60 St. George St., Toronto, Ontario M5S 1A7,
Canada\\
$^{(2)}$Chemical Physics Theory Group, Chemistry Department, and Center for
Quantum Information and Quantum Control, University of Toronto, 80 St.
George St., Toronto, Ontario M5S 3H6, Canada}

\begin{abstract}
We introduce a generalized theory of decoherence-free subspaces and
subsystems (DFSs), which do not require accurate initialization. We
derive a new set of conditions for the existence of DFSs within this
generalized framework. By relaxing the initialization requirement we
show that a DFS can tolerate arbitrarily large preparation errors.
This has potentially significant implications for experiments
involving DFSs, in particular for the experimental implementation,
over DFSs, of the large class of quantum algorithms which can
function with arbitrary input states.
\end{abstract}

\maketitle

\section{Introduction}

In recent years much effort has been expended to develop methods for
tackling the deleterious interaction of controlled quantum systems
with their environment. This effort has been motivated in large part
by the need to overcome decoherence in quantum information processing
tasks, a goal which was thought to be unattainable at first
\cite{Landauer:95,Unruh:95,Haroche:96}. Decoherence-free (or
noiseless) subspaces \cite{Duan:98,Zanardi:97c,Lidar:PRL98,Lidar:00a}
and subsystems \cite{Knill:99a,DeFilippo:00,Kempe:00,Yang:01} (DFSs)
are among the methods which have been proposed to this end, and also
experimentally realized in a variety of systems
\cite{Kwiat:00,Kielpinski:01,Fortunato:01,Viola:01b}. In this manner
of passive quantum error correction, one uses symmetries in the form
of the interaction between system and environment to find a
\textquotedblleft quiet corner\textquotedblright\ in the system
Hilbert space not experiencing this interaction. Of the various
methods of quantum error correction, so far only DFSs have been
combined with quantum algorithms in the presence of decoherence
\cite{Mohseni:02,Ollerenshaw:02}.  For a review of DFSs and a
comprehensive list of references see Ref.~\cite{LidarWhaley:03}.

We have re-examined the theoretical foundation of DFSs and have found that
the conditions for their existence can be generalized. It is our purpose in
this paper to present these generalized conditions. Our most significant
result is a drastic relaxation of the initialization condition for DFSs:
whereas it was previously believed that one must be able to perfectly
initialize a state inside a DFS, here we show that this does in fact need
not be so. Instead one can tolerate an arbitrarily large preparation error,
which in turn means significantly relaxed experimental preparation
conditions. In contrast, only a small preparation error can be tolerated
when quantum error correcting codes (QECC) are used to overcome decoherence
\cite{Preskill:99}. Whether a similar generalization is possible in the case
of QECC is an interesting open question, the answer to which may be
within the realm of very recent results strengthening the DFS/QECC
connection \cite{Kribs:05}.

The relaxation of the initialization requirement is perhaps most significant
in light of a series of results showing that a class of important quantum
algorithms (Shor \cite{Shor:97}, Grover \cite{Grover:96}, and
Deutsch-Josza \cite{Deutsch:92} included) can be successfully executed under
imperfect initialization conditions
\cite{BBBGL98,Lidar:PRA99Grover,Lidar:PRA01Grover,Biham:02,Parker:00,Knill:98a,Chi:01,Kim:02,Chi:05}.
This means that imperfectly initialized DFSs can be used as a
``substrate'' for running these algorithms.

To present our results we first review and re-examine the previous results
on DFSs, in Section~\ref{review}. We do so for both general completely
positive (CP) maps and for Markovian dynamics. The definitions we give for
DFSs in these two cases are slightly different, reflecting the fact that
Markovian dynamics is always continuous in time, whereas CP maps can also
describe discrete-time evolution. In Section~\ref{generalized}, we present
our generalized DFS conditions for CP maps and for Markovian dynamics. We
illustrate the new conditions for Markovian dynamics with an example which
reveals some of the new features. In Section~\ref{algo} we discuss the
implications of our relaxed initialization condition in the context of
quantum algorithms. Section \ref{ME} is devoted to a case-study of
non-Markovian dynamics, intermediate between (formally exact) CP maps and
(approximate) Markovian dynamics. A unique formulation does not exist in
this case, and we consider the master equation introduced in Ref.~\cite{ShabaniLidar:05}. The analytical solvability of this equation permits a
rigorous derivation of the conditions for a DFS. For clarity of presentation
we defer most supporting calculations to the appendices.

\section{Review of Previous Conditions for Decoherence-Free Subspaces and
Subsystems}

\label{review}

We refer the reader to Ref.~\cite{LidarWhaley:03} for a detailed review,
including many references and historical context. Here we focus on aspects
of direct relevance to our new results.

\subsection{Decoherence-Free Subspaces}

Consider a system with Hilbert space $\mathcal{H}_{S}$. In
Refs.~\cite{Zanardi:97c,Lidar:PRL98,Zanardi:97a,Lidar:PRL99,Lidar:00a} a
subspace $\mathcal{H}_{\mathrm{DFS}}\subset \mathcal{H}_{S}$ was
called decoherence-free if any state $\rho _{S}(0)$ of the system
initially prepared in this subspace is unitarily related to the final
state $\rho _{S}(t)$ of the system, i.e.,
\begin{equation}
\rho _{S}(0)=\mathcal{P}_{\mathrm{d}}\rho _{S}(0)\mathcal{P}_{\mathrm{d}
}\Longrightarrow \rho _{S}(t)=\mathbf{U}\rho _{S}(0)\mathbf{U}^{\dagger }.
\label{DFS-Def}
\end{equation}
Here $\mathbf{U}$ is unitary and $\mathcal{P}_{\mathrm{d}}$ is the
projection operator onto $\mathcal{H}_{\mathrm{DFS}}$. Important and
motivating early examples of DFSs were given in
\cite{Palma:96,Duan:97PRL,Duan:98,Zanardi:98}. An alternative
definition of a DFS is as a subspace in which the state purity is
always one \cite{ZanardiLidar:04}; here we will not pursue this
approach.

To exploit DF-states for quantum information preservation one needs a
method to experimentally verify these states \cite{Viola:03}, but from
a theoretical standpoint one needs to first formulate the effect of
the environment. In the following, we consider general CP maps and
Markovian dynamics.

\subsubsection{Completely Positive Maps}

The modeling of environmental effects on an open quantum system has been a
challenging problem since at least the 1950's \cite{Nakajima:58,Zwanzig:60a}%
, but under certain simplifying assumptions one can obtain a simple form for
the dynamical equations of open systems \cite{Breuer:book}. For example, the
assumption of an initially decoupled state of system and bath, $\rho
_{SB}(0)=\rho _{S}(0)\otimes \rho _{B}$, results in a CP map known as the
Kraus operator sum representation \cite{Kraus:83}:%
\begin{eqnarray}
\rho _{S}(t) &=& \mathrm{Tr}_{B}[\mathbf{\Lambda} (t)\left( \rho
_{S}(0)\otimes \rho_{B}\right) \mathbf{\Lambda} (t)^{\dag }]  \notag \\
&=& \sum_{\alpha }\mathbf{E}_{\alpha }(t)\rho
_{S}(0)\mathbf{E}_{\alpha }^{\dagger }(t).  \label{Kraus}
\end{eqnarray}%
Here
\begin{eqnarray}
\mathbf{\Lambda} (t)=\mathcal{T}\exp (-i\int_{0}^{t}\mathbf{H}(s)ds)
\end{eqnarray}%
is the unitary propagator for the joint evolution of system and bath
governed by total Hamiltonian $\mathbf{H}$ ($\mathcal{T}$ denotes
time-ordering and we work in units such that $\hbar =1$); the
\textquotedblleft Kraus operators\textquotedblright\ $\{\mathbf{E}_{\alpha
}\}$ are given by
\begin{eqnarray}
\mathbf{E}_{\alpha }=\sqrt{\lambda _{\nu }}\langle \mu |\mathbf{\Lambda }%
|\nu \rangle ;\text{ \ \ \ \ \ \ \ }\alpha =(\mu ,\nu ),  \label{operators}
\end{eqnarray}%
where $|\mu \rangle ,|\nu \rangle $ are bath states in the spectral
decomposition $\rho _{B}=\sum_{\nu }\lambda _{\nu }|\nu \rangle \langle \nu
| $. Trace preservation of $\rho _{S}(t)$ implies the sum rule
\begin{eqnarray}
\sum_{\alpha }\mathbf{E}_{\alpha }^{\dagger }\mathbf{E}_{\alpha }=\mathbf{I}%
_{S},
\end{eqnarray}%
where $\mathbf{I}_{S}$ is the identity operator on the system.

In \cite{Lidar:PRL99} a DFS-condition was derived for general CP maps of
this type. We denote the subspace of states orthogonal to $\mathcal{H}_{%
\mathrm{DFS}}$ by $\mathcal{H}_{\mathrm{DFS}^{\bot }}$, so that $\mathcal{H}%
_{S}=\mathcal{H}_{\mathrm{DFS}}\oplus \mathcal{H}_{\mathrm{DFS}^{\bot }}$.
According to Eq.~(4) in \cite{Lidar:PRL99} the Kraus operators take the
block-diagonal form%
\begin{eqnarray}
\mathbf{E}_{\alpha }=\left(
\begin{array}{cc}
c_{\alpha }\mathbf{U}_{\mathrm{DFS}} & \mathbf{0} \\
\mathbf{0} & \mathbf{B}_{\alpha }%
\end{array}%
\right) ,  \label{OLD-KRAUS-DFS}
\end{eqnarray}%
where the upper (lower) non-zero block acts entirely inside $\mathcal{H}_{%
\mathrm{DFS}}$ ($\mathcal{H}_{\mathrm{DFS}^{\bot }}$); $\mathbf{U}_{\mathrm{%
DFS}}$ is a unitary matrix that is independent of the Kraus operator label $%
\alpha $; $c_{\alpha }$ is a scalar ($\sum_{\alpha }|c_{\alpha }|^{2}=1$);
and $\mathbf{B}_{\alpha }$ is arbitrary, except that $\sum_{\alpha }\mathbf{B%
}_{\alpha }^{\dagger }\mathbf{B}_{\alpha }=\mathbf{I}_{\mathrm{DFS}^{\bot }}$%
. It is simple to verify that the DFS definition (\ref{DFS-Def}) is
satisfied in this case, with $\mathbf{U}=\mathbf{U}_{\mathrm{DFS}}$.

Theorem~1 in \cite{Lidar:PRL99} reads: \textquotedblleft A subspace $%
\mathcal{H}_{\mathrm{DFS}}$ is a DFS iff all Kraus operators have an
identical unitary representation upon restriction to it, up to a
multiplicative constant.\textquotedblright\ This theorem is actually
compatible with a more general form for the Kraus operators than Eq.~(\ref{OLD-KRAUS-DFS}), since \textquotedblleft upon restriction to
it\textquotedblright\ concerns only the upper-left block of $\mathbf{E}_{\alpha }$. We derive
the most general form of $\mathbf{E}_{\alpha }$ in Section \ref{generalized}
below, and find that, indeed, a more general form than Eq.~(\ref{OLD-KRAUS-DFS}) is possible: one of the off-diagonal blocks need not
vanish. In other words, leakage from $\mathcal{H}_{\mathrm{DFS}^{\bot }}$
into $\mathcal{H}_{\mathrm{DFS}}$ is permitted. As we further show in
Section \ref{generalized}, the form (\ref{OLD-KRAUS-DFS}) in fact appears in
the context of unital channels.

\subsubsection{Markovian Dynamics}

The most general form of CP Markovian dynamics is given by the Lindblad
equation \cite{Gorini:76,Lindblad:76,Alicki:87}:
\begin{eqnarray}
\frac{\partial \rho _{S}}{\partial t} &=&-i[\mathbf{H}_{S},\rho _{S}]+%
\mathcal{L}[\rho _{S}],  \notag \\
\mathcal{L}\cdot &=&\sum_{\alpha }\mathbf{F}_{\alpha }\cdot \mathbf{F}%
_{\alpha }^{\dag }-\frac{1}{2}\mathbf{F}_{\alpha }^{\dag }\mathbf{F}_{\alpha
}\cdot -\frac{1}{2}\cdot \mathbf{F}_{\alpha }^{\dag }\mathbf{F}_{\alpha },
\label{Lindblad}
\end{eqnarray}%
where $\mathbf{F}_{\alpha }$ are bounded (or unbounded, if subject to
appropriate domain restrictions \cite{Davies:77,LidarShabaniAlicki:05})
operators acting on $\mathcal{H}_{\mathrm{S}}$, and where $\mathbf{H}_{S}$
may include a Lamb shift \cite{Lidar:CP01}. Given such dynamics, one
restores unitarity [i.e., the DFS definition (\ref{DFS-Def}) with $\mathbf{U}
$ generated by the Hamiltonian $\mathbf{H}_{S}$] if the Lindblad term $%
\mathcal{L}[\rho _{S}]$ can be eliminated. According to Refs.~\cite%
{Lidar:PRL98,Zanardi:98a}, a necessary and sufficient condition for this to
be the case is
\begin{eqnarray}
\mathbf{F}_{\alpha }|i\rangle =c_{\alpha }|i\rangle ,  \label{Fa}
\end{eqnarray}
where $\mathcal{H}_{\mathrm{DFS}}=\mathrm{Span}\{|i\rangle \}$ and $%
\{c_{\alpha }\}$ are arbitrary complex scalars. Thus the Lindblad operators
can be written in block-form as follows:%
\begin{eqnarray}
\mathbf{F}_{\alpha }=\left(
\begin{array}{cc}
c_{\alpha }\mathbf{I} & \mathbf{A}_{\alpha } \\
\mathbf{0} & \mathbf{B}_{\alpha }%
\end{array}%
\right) ,  \label{Markov-DFS}
\end{eqnarray}%
with the blocks on the diagonal corresponding once again to operators
restricted to $\mathcal{H}_{\mathrm{DFS}}$ and $\mathcal{H}_{\mathrm{DFS}%
^{\bot }}$. Note the appearance of the off-diagonal block $\mathbf{A}%
_{\alpha }$ mixing $\mathcal{H}_{\mathrm{DFS}}$ and $\mathcal{H}_{\mathrm{DFS%
}^{\bot }}$; its presence is permitted since the DFS condition (\ref{Fa})
gives no information about matrix elements of the form $\langle i|\mathbf{F}%
_{\alpha }|j^{\bot }\rangle $, with $|i\rangle \in \mathcal{H}_{\mathrm{DFS}%
} $ and $|j^{\bot }\rangle \in \mathcal{H}_{\mathrm{DFS}^{\bot }}$.

As observed in Refs.~\cite{Lidar:PRL98,Lidar:PRL99}, one should in addition
require that $\mathbf{H}_{S}$ does not mix DF states with non-DF ones, i.e.,
mixed matrix elements of the type $\langle j^{\bot }|\mathbf{H}_{S}|i\rangle
$, with $|i\rangle \in \mathcal{H}_{\mathrm{DFS}}$ and $|j^{\bot }\rangle
\in \mathcal{H}_{\mathrm{DFS}^{\bot }}$, should vanish. We show below that
this condition must be made more stringent.

\subsection{Noiseless Subsystems}

An important observation made in Ref.~\cite{Knill:99a} is that there is no need to restrict the
decoherence-free dynamics to a subspace. A more general situation is when
the DF dynamics is a \textquotedblleft subsystem\textquotedblright , or a
factor in a tensor product decomposition of subspace. Following Ref.~\cite{Knill:99a}, this comes about as follows. Consider the dynamics
of a system $S$ coupled to a bath $B$ via the Hamiltonian
\begin{eqnarray}
\mathbf{H}=\mathbf{H}_{S}\otimes \mathbf{I}_{B}+\mathbf{I}_{S}\otimes
\mathbf{H}_{B}+\mathbf{H}_{I},
\end{eqnarray}%
where $\mathbf{H}_{S}$ ($\mathbf{H}_{B}$), the system (bath) Hamiltonian,
acts on the system (bath) Hilbert space $\mathcal{H}_{S}$ ($\mathcal{H}_{B}$); $\mathbf{I}_{S}$ ($\mathbf{I}_{B}$) is the identity operator on the
system (bath) Hilbert space; $\mathbf{H}_{I}$ is the interaction term of
Hamiltonian which can be written in general as $\sum_{\alpha }\mathbf{S}%
_{\alpha }\otimes \mathbf{B}_{\alpha }$. If the system Hamiltonian $\mathbf{H%
}_{S}$ and the system components of the interaction Hamiltonian, the $%
\mathbf{S}_{\alpha }$'s, form an algebra $\mathcal{S}$, it must be $\dagger $%
-closed to preserve the unitarity of system-bath dynamics. Now, if $\mathcal{%
A}$ is a $\dagger $-closed operator algebra which includes the identity
operator, then a fundamental theorem of C$^{\ast }$ algebras states that $%
\mathcal{A}$ is a reducible subalgebra of the full algebra of operators \cite%
{Landsman:98a}. This theorem implies that the algebra is isomorphic to a
direct sum of $d_{J}\times d_{J}$ complex matrix algebras, each with
multiplicity $n_{J} $:%
\begin{eqnarray}
\mathcal{S}\cong \bigoplus_{J\in \mathcal{J}}\mathbf{I}_{n_{J}}\otimes
\mathcal{M}(d_{J},\mathbb{C})  \label{dfss-algebra}
\end{eqnarray}%
Here $\mathcal{J}$ is a finite set labeling the irreducible components of $%
\mathcal{S}$, and $\mathcal{M}(d_{J},\mathbb{C})$ denotes a $d_{J}\times
d_{J}$ complex matrix algebra. Associated with this decomposition of the
algebra $\mathcal{S}$ is a decomposition of the system Hilbert space:
\begin{eqnarray}
\mathcal{H}_{S}=\bigoplus_{J\in \mathcal{J}}\mathbb{C}^{n_{J}}\otimes
\mathbb{C}^{d_{J}}.  \label{dfss-decomposition}
\end{eqnarray}%
If we encode quantum information into a subsystem (factor) $\mathbb{C}%
^{n_{J}}$ it is preserved, since the noise algebra $\mathcal{S}$ acts
trivially (as $\mathbf{I}_{n_{J}}$). In such a case $\mathbb{C}^{n_{J}}$ is
called a decoherence-free, or noiseless subsystem (NS) \cite{Knill:99a}.
Examples of this construction were given independently in Refs.~\cite{DeFilippo:00,Yang:01}.

\subsubsection{Completely Positive Maps}

As the Kraus operators are given by Eq.~(\ref{operators}), they take the
form of the decomposition (\ref{dfss-algebra}):%
\begin{eqnarray}
\mathbf{E}_{\alpha }=\bigoplus_{J\in \mathcal{J}}\mathbf{I}_{n_{J}}\otimes
\mathbf{M}_{\alpha }(d_{J}),  \label{Kraus-DFSS}
\end{eqnarray}%
where $\mathbf{M}_{\alpha }(d_{J})$ is an arbitrary $d_{J}$-dimensional
complex matrix. Therefore a factor $\mathbb{C}^{n_{J}}$ is a NS if the Kraus
operators have the representation (\ref{Kraus-DFSS}).

\subsubsection{Markovian Dynamics}

The aforementioned reducibility theorem \cite{Landsman:98a} does not
apply directly in the Markovian case, since the set of Lindblad
operators $\{\mathbf{F}_{\alpha }\}$ need not be closed under
conjugation. Nevertheless, as shown in \cite{Kempe:00}, the concept of
a subsystem applies in the Markovian case as well: the condition for a
NS was found to be
\begin{eqnarray}
\mathbf{F}_{\alpha }\mathcal{P}_{\mathrm{d}}=\mathbf{I}_{n_{J}}\otimes
\mathbf{M}_{\alpha }(d_{J})\mathcal{P}_{\mathrm{d}},  \label{Markovian-DFSS}
\end{eqnarray}

with the $\mathbf{M}_{\alpha }$ again being arbitrary complex matrices
and $\mathcal{P}_{\mathrm{d}}$ being the projection operator onto a
given subspace $\mathbb{C}^{n_{J}}\otimes \mathbb{C}^{d_{J}}$. The NS
is then a factor $\mathbb{C}^{n_{J}}$ as in
Eq.~(\ref{dfss-decomposition}), with the same tensor product structure
as in Eq.~(\ref{Markovian-DFSS}).

\section{Generalized Conditions for Decoherence-Free Subspaces and Subsystems}

\label{generalized}

We now proceed to re-examine the conditions for the existence of
decoherence-free subspaces and subsystems. We will show that the conditions
presented in the papers laying the general theoretical foundation
\cite{Zanardi:97c,Lidar:PRL98,Knill:99a,Zanardi:97a,Lidar:PRL99,Zanardi:98a,Kempe:00}
, can be generalized and sharpened, both for CP maps and for Markovian
dynamics. Our main new finding is that the preparation step can tolerate
arbitrarily large errors. Relatedly, we consider the possibility of leakage
from outside of the protected subspace/subsystem into it. Previous studies
did not allow for this possibility, but we will show that it can be
permitted under appropriate restrictions. In doing so we generalize the
definition of a NS with respect to the original definition that relied on
the algebraic isomorphism (\ref{dfss-algebra}) (see Ref.~\cite{Kribs:05} for
a related recent result). In the case of Markovian
dynamics, our main new finding is that if one demands perfect initialization
into a DFS then the condition on the Hamiltonian component of the evolution
is modified compared to previous studies.

The derivation of these results is somewhat tedious. Hence, for clarity of
presentation we focus on presenting our generalized conditions in this
section. Mathematical proofs are deferred to the appendices. We begin with
the simpler case of decoherence-free subspaces and consider the case of CP
maps and Markovian dynamics. We then move on to the case of decoherence-free
(noiseless) subsystems. The case of non-Markovian continuous-time dynamics
is treated later, in Section~\ref{ME}.

\subsection{Decoherence-Free Subspaces}

The system density matrix $\rho _{S}$ is an operator on the entire system
Hilbert space $\mathcal{H}_{S}$, which we assume to be decomposable into a
direct sum as $\mathcal{H}=\mathcal{H}_{\mathrm{DFS}}\mathbf{\oplus }%
\mathcal{H}_{\mathrm{DFS}^{\bot }}$. It is convenient for our purposes to
represent the system state (and later on the Kraus and Lindblad operators)
in a matrix form whose block structure corresponds to this decomposition of
the Hilbert space. Thus the system density matrix takes the form
\begin{equation}
\rho _{S}=\left(
\begin{array}{cc}
\rho _{\mathrm{DFS}} & \rho _{2} \\
\rho _{2}^{\dag } & \rho _{3}%
\end{array}%
\right) ,
\label{eq:rho-DFS-blocks}
\end{equation}%
We also define a projector
\begin{equation}
\mathcal{P}_{\mathrm{DFS}}=\left(
\begin{array}{cc}
\mathbf{I}_{\mathrm{DFS}} & \mathbf{0}%
\end{array}%
\right) ,
\end{equation}%
so that $\rho _{\mathrm{DFS}}=\mathcal{P}_{\mathrm{DFS}}\rho _{S}\mathcal{P}%
_{\mathrm{DFS}}^{\dag }$. Finally,
\begin{equation}
\mathcal{P}_{\mathrm{d}}=\left(
\begin{array}{cc}
\mathbf{I}_{\mathrm{DFS}} & \mathbf{0} \\
\mathbf{0} & \mathbf{0}%
\end{array}%
\right) , \quad
\mathcal{P}_{\mathrm{d}^\bot}=\left(
\begin{array}{cc}
\mathbf{0} & \mathbf{0} \\
\mathbf{0} & \mathbf{I}_{\mathrm{DFS}}%
\end{array}%
\right)
\end{equation}
are projection operators onto $\mathcal{H}_{\mathrm{DFS}}$ and
$\mathcal{H}_{\mathrm{DFS}^\bot}$, respectively.

\subsubsection{Completely Positive Maps}

The original concept of a DFS, Eq.~(\ref{DFS-Def}), poses a practical
problem: the perfect initialization of a quantum system inside a DFS might
be challenging in many cases. Therefore we introduce a generalized
definition to relax this constraint:

\begin{mydefinition}
Let the system Hilbert space $\mathcal{H}_S$ decompose into a direct sum as $%
\mathcal{H}=\mathcal{H}_{\mathrm{DFS}}\mathbf{\oplus }\mathcal{H}_{\mathrm{%
DFS}^{\bot }}$, and partition the system state $\rho _{S}$ accordingly into
blocks, as in Eq.~(\ref{eq:rho-DFS-blocks}). Assume $\rho _{\mathrm{DFS}}(0)=%
\mathcal{P}_{\mathrm{DFS}}\rho _{S}(0)\mathcal{P}_{\mathrm{DFS}}^{\dag }\neq
\mathbf{0}$. Then $\mathcal{H}_{\mathrm{DFS}}$ is called decoherence-free
iff the initial and final DFS-blocks of $\rho _{S}$ are unitarily related:
\begin{equation}
\rho _{\mathrm{DFS}}(t)=\mathbf{U}_{\mathrm{DFS}}\rho _{\mathrm{DFS}}(0)%
\mathbf{U}_{\mathrm{DFS}}^{\dag },  \label{eq:newDFSdef}
\end{equation}%
where $\mathbf{U}_{\mathrm{DFS}}$ is a unitary matrix acting on $\mathcal{H}%
_{\mathrm{DFS}}$.
\end{mydefinition}

\begin{mydefinition}
  \label{def:per}
Perfect initialization (DF subspaces): $\rho_2 =\mathbf{0}$ and $\rho_3 = \mathbf{0}$ in Eq.~(%
\ref{eq:rho-DFS-blocks}).
\end{mydefinition}

\begin{mydefinition}
  \label{def:imp}
Imperfect initialization (DF subspaces): $\rho_2$ and/or $\rho_3$ in Eq.~(%
\ref{eq:rho-DFS-blocks}) are non-vanishing.
\end{mydefinition}

We prove in Appendix~\ref{app1}:

\begin{mytheorem}
\label{th:CP-DFS-imperfect}Assume imperfect initialization. Let $\mathbf{U}$
be unitary, $c_{\alpha }$ scalars satisfying $\sum_{\alpha }|c_{\alpha
}|^{2}=1$, and $\mathbf{B}_{\alpha }$ arbitrary operators on $\mathcal{H}_{%
\mathrm{DFS}^{\bot }}$ satisfying $\sum_{\alpha }\mathbf{B}_{\alpha
}^{\dagger }\mathbf{B}_{\alpha }=\mathbf{I}_{\mathrm{DFS}^{\bot }}$. A
necessary and sufficient condition for the existence of a DFS with respect
to CP maps is that the Kraus operators have a matrix representation of the
form%
\begin{eqnarray}
\mathbf{E}_{\alpha }=\left(
\begin{array}{cc}
c_{\alpha }\mathbf{U} & \mathbf{0} \\
\mathbf{0} & \mathbf{B}_{\alpha }%
\end{array}%
\right) .  \label{Mat-Rep}
\end{eqnarray}
\end{mytheorem}

This form is identical to the previous result (\ref{OLD-KRAUS-DFS}), with
the important distinction that due to the new definition of a DFS, Eq.~(\ref%
{eq:newDFSdef}), the theorem holds not just for states initialized perfectly
into $\mathcal{H}_{\mathrm{DFS}}$, but for arbitrary initial states. Note
that unlike fault-tolerant QECC, where the initial state must be
sufficiently close to a valid code state \cite{Preskill:99}, here the
initial state can be arbitrarily far from a DFS-code state, as long as the
initial projection into the DFS is non-vanishing.

These observations lead us to reconsider the original definition, wherein
the system \emph{is} initialized inside the DFS. This situation admits more
general Kraus operators. Specifically, we prove Appendix~\ref{app1} that:

\begin{mycorollary}
\label{DFS} Assume perfect initialization. Then the DFS condition is:%
\begin{eqnarray}
\mathbf{E}_{\alpha }=\left(
\begin{array}{cc}
c_{\alpha }\mathbf{U} & \mathbf{A}_{\alpha } \\
\mathbf{0} & \mathbf{B}_{\alpha }%
\end{array}%
\right) ,  \label{Mat-Rep2}
\end{eqnarray}%
where $\mathbf{U}$ is unitary.
\end{mycorollary}
Note that due to the sum rule $\sum_{\alpha }\mathbf{E}_{\alpha
}^{\dagger }\mathbf{E}_{\alpha }=\mathbf{I}$ the otherwise arbitrary
operators $\mathbf{A}_{\alpha }$ and $\mathbf{B}_{\alpha }$ satisfy the
constraints (i) $\sum_{\alpha }\mathbf{A}_{\alpha }^{\dagger }\mathbf{A}%
_{\alpha }+\mathbf{B}_{\alpha }^{\dagger }\mathbf{B}_{\alpha }=\mathbf{I}_{%
\mathrm{DFS}^{\bot }}$ and (ii) $\sum_{\alpha }c_{\alpha }^{\ast }\mathbf{A}%
_{\alpha }=\mathbf{0}$, and\ where additionally the scalars $c_{\alpha }$
satisfy (iii) $\sum_{\alpha }|c_{\alpha }|^{2}=1$.

In contrast to the diagonal form in the previous conditions (\ref%
{OLD-KRAUS-DFS}) and (\ref{Mat-Rep}), Eq.~(\ref{Mat-Rep2}) allows for the
existence of the off-diagonal term $\mathbf{A}_{\alpha }$, which permits
leakage from $\mathcal{H}_{\mathrm{DFS}^{\bot }}$ into $\mathcal{H}_{\mathrm{%
DFS}}$. This more general form of the Kraus operators imply that a larger
class of noise processes allow for the existence of DFSs, as compared to the
previous condition (\ref{OLD-KRAUS-DFS}).\footnote{We re-emphasize that Theorem 1 in \cite{Lidar:PRL99} is
compatible with Eq.~(\ref{Mat-Rep2}); the latter generalizes the explicit
matrix representation Eq.~(4) given in that paper [condition (\ref%
{OLD-KRAUS-DFS}) in the present paper], but does not invalidate Theorem 1 in
\cite{Lidar:PRL99}.}

\subsubsection{Unital Maps}

A unital (sometimes called bi-stochastic) channel is a CP map $\mathbf{\Phi }%
(\rho )=\sum_{\alpha }\mathbf{E}_{\alpha }\rho \mathbf{E}_{\alpha }^{\dagger
}$ that preserves the identity operator: $\mathbf{\Phi }(\mathbf{I}%
)=\sum_{\alpha }\mathbf{E}_{\alpha }\mathbf{E}_{\alpha }^{\dagger }=\mathbf{I%
}$. Consider the fixed points of $\mathbf{\Phi }$, i.e., $\mathrm{Fix}(%
\mathbf{\Phi })\equiv \{\rho :\mathbf{\Phi }(\rho )=\rho \}$. Such states,
which are invariant under $\mathbf{\Phi }$, are clearly examples of
DF-states of the corresponding channel.

Recently it has been shown that the fixed point set of unital CP maps is the
commutant of the algebra generated by Kraus operators \cite{Kribs:03}. In
other words, if $\mathcal{E}$ is the set of all polynomials in $\{\mathbf{E}%
_{\alpha }\}$, or ${\mathcal{E}}=\mathrm{Alg}\{\mathbf{E}_{\alpha }\}$, then
\begin{eqnarray}
\mathrm{Fix}(\mathbf{\Phi })=\{\mathbf{T}\in {\mathcal{B}}({\mathcal{H}}):[%
\mathbf{T},{\mathcal{E}}]=\mathbf{0}\},
\end{eqnarray}%
where ${\mathcal{B}}({\mathcal{H}})$ is the (Banach)\ space of all bounded
operators on the Hilbert space ${\mathcal{H}}$. In other words, the fixed
points of a unital CP map, which are DF states, can alternatively be
characterized as the commutant of $\mathrm{Alg}\{\mathbf{E}_{\alpha }\}$,
i.e., the set $\{\mathbf{T}\}$. It is our purpose in this subsection to show
that, under our generalized definition of DFSs, this characterization of DF states is sufficient but not necessary.

Consider the generalized DFS-condition (\ref{Mat-Rep2}) applied to unital
maps. We have

\begin{eqnarray}
\mathbf{\Phi }(\rho )=\sum_{\alpha }\left(
\begin{array}{cc}
c_{\alpha }\mathbf{I}_{\mathrm{DFS}} & \mathbf{A}_{\alpha } \\
\mathbf{0} & \mathbf{B}_{\alpha }%
\end{array}%
\right) \rho \left(
\begin{array}{cc}
c_{\alpha }^{\ast }\mathbf{I}_{\mathrm{DFS}} & \mathbf{0} \\
\mathbf{A}_{\alpha }^{\dagger } & \mathbf{B}_{\alpha }^{\dagger }%
\end{array}%
\right) .
\end{eqnarray}%
Unitality, $\mathbf{\Phi }(\mathbf{I})=\mathbf{I}$, together with $%
\sum_{\alpha }|c_{\alpha }|^{2}=1$ implies:

\begin{eqnarray}
\left(
\begin{array}{cc}
\mathbf{I}_{\mathrm{DFS}}+\sum_{\alpha }\mathbf{A}_{\alpha }\mathbf{A}%
_{\alpha }^{\dagger } & \sum_{\alpha }\mathbf{A}_{\alpha }\mathbf{B}_{\alpha
}^{\dagger } \\
\sum_{\alpha }\mathbf{B}_{\alpha }\mathbf{A}_{\alpha }^{\dagger } &
\sum_{\alpha }\mathbf{B}_{\alpha }\mathbf{B}_{\alpha }^{\dagger }%
\end{array}%
\right) =\mathbf{I}.
\end{eqnarray}%
This implies the vanishing of the matrices $\mathbf{A}_{\alpha }$, so that
we are left with the Kraus operators in the simple block-diagonal form:

\begin{eqnarray}
\mathbf{E}_{\alpha }=\left(
\begin{array}{cc}
c_{\alpha }\mathbf{I} & \mathbf{0} \\
\mathbf{0} & \mathbf{B}_{\alpha }%
\end{array}%
\right) ,  \label{unitalE}
\end{eqnarray}%
together with the additional constraint $\sum_{\alpha }\mathbf{B}_{\alpha }%
\mathbf{B}_{\alpha }^{\dagger }=\mathbf{I}_{\mathrm{DFS}^{\bot }}$ (which,
in the present unital case, naturally supplements the previously derived
normalization constraint $\sum_{\alpha }\mathbf{B}_{\alpha }^{\dag }\mathbf{B%
}_{\alpha }=\mathbf{I}_{\mathrm{DFS}^{\bot }}$). Thus, unitality restricts
the class of Kraus operators, so that in fact we must assume the
DFS-condition (\ref{Mat-Rep}) rather than (\ref{Mat-Rep2}). This then means
that we may consider the generalized DFS definition Eq.~(\ref{eq:newDFSdef}).

Next, let us find the commutant of this class of Kraus operators. First,
\begin{eqnarray}
\mathrm{Alg}\{\mathbf{E}_{\alpha }\}=\{\left(
\begin{array}{cc}
\mathrm{poly}(c_{\alpha })\mathbf{I} & \mathbf{0} \\
\mathbf{0} & \mathrm{poly}(\mathbf{B}_{\alpha })%
\end{array}%
\right) \},
\end{eqnarray}%
where $\mathrm{poly}(x)$ denotes all possible polynomials in $x$.
Representing an arbitrary operator $\mathbf{T}\in {\mathcal{B}}({\mathcal{H}}%
)$ in the form
\begin{eqnarray}
\mathbf{T}=\left(
\begin{array}{cc}
\mathbf{L} & \mathbf{M} \\
\mathbf{N} & \mathbf{P}%
\end{array}%
\right) ,
\end{eqnarray}%
it is simple to derive that the commutant of $\mathrm{Alg}\{\mathbf{E}%
_{\alpha }\}$ is the space of matrices $\mathbf{T}$ of the form
\begin{eqnarray}
\mathbf{T}=\left(
\begin{array}{cc}
\mathbf{L} & \mathbf{0} \\
\mathbf{0} & c\mathbf{I}%
\end{array}%
\right) ,  \label{eq:T}
\end{eqnarray}%
where $\mathbf{L}$ and $c$ are arbitrary. The aforementioned theorem \cite%
{Kribs:03} states that the fixed-point set of the channel, i.e., the DF
states, coincides with this commutant. Of course, for $\mathbf{T}$ to be a
proper quantum state it must be Hermitian and have unit trace, whence $c\geq
0$ and $\mathbf{L}$ is Hermitian. Subject to these constraints we see that
the aforementioned theorem \cite{Kribs:03} gives a sufficient, but not
necessary characterization of the allowed DF states. Indeed, the form (\ref%
{eq:T}) arises as a special case of our considerations, where we allow for $%
\mathbf{T}$ to be a state with support in $\mathcal{H}_{\mathrm{DFS}^{\bot
}} $, but not of the most general form allowed by Eq.~(\ref{eq:newDFSdef}),
which includes off-diagonal blocks.

\subsubsection{Markovian Dynamics}

In the case of CP maps we are only interested in the output state and the
intermediate-time states are ignored. Since, as is well known, Markovian
dynamics is a special case of CP maps (e.g., \cite{Alicki:87,Lidar:CP01}),
one may of course apply the results we have obtained above for general CP
maps in the Markovian case as well, provided one is only interested in the
state at the end of the Markovian channel. However, one may instead be
interested in a different notion of decoherence-freeness, wherein the system
remains DF throughout the entire evolution. Such a notion is more suited to
experiments in which the final time is not a priori known. This is the
notion we will pursue here in our treatment of continuous-time dynamics, in
both the Markovian and non-Markovian cases. Thus, while we allow that the
system not be fully initialized into the DFS, we require that the component
that is, undergoes unitary dynamics \emph{at all times}. Correspondingly, we
define a DFS in the Markovian case as follows:

\begin{mydefinition}
Let the system Hilbert space $\mathcal{H}_S$ decompose into a direct sum as $%
\mathcal{H}_S=\mathcal{H}_{\mathrm{DFS}}\mathbf{\oplus }\mathcal{H}_{\mathrm{%
DFS}^{\bot }}$, and partition the system state $\rho _{S}$ accordingly into
blocks. Let $\mathcal{P}_{\mathrm{DFS}}$ be a projector onto $\mathcal{H}_{%
\mathrm{DFS}}$ and assume $\rho _{\mathrm{DFS}}(0)\equiv \mathcal{P}_{%
\mathrm{DFS}}\rho _{S}(0)\mathcal{P}_{\mathrm{DFS}}^{\dagger }\neq \mathbf{0}$. Then $%
\mathcal{H}_{\mathrm{DFS}}$ is called decoherence-free iff $\rho _{\mathrm{%
DFS}}$ undergoes Schr\"{o}dinger-like dynamics,
\begin{equation}
\frac{\partial \rho _{\mathrm{DFS}}}{\partial
  t}=-i[\mathbf{H}_{\mathrm{DFS}},\rho _{\mathrm{DFS}}],
\label{DFS-Lindblad}
\end{equation}
where $\mathbf{H}_{\mathrm{DFS}}$ is a Hermitian operator.
\end{mydefinition}

Before presenting the DFS conditions, let us recall the quantum trajectories
interpretation of Markovian dynamics \cite{Dalibard:92,Gisin:92,Plenio:98}.
Expanding Eq.~(\ref{Lindblad}) to first order in the short time-interval $%
\tau $ yields the CP map
\begin{eqnarray}
\rho _{S}(t+\tau )=\sum_{\beta =0}\mathbf{W}_{\beta }\rho (t)\mathbf{W}%
_{\beta }^{\dagger },
\end{eqnarray}%
where
\begin{eqnarray}
\mathbf{W}_{0} &=&\mathbf{I}-i\tau \mathbf{H}_{S}-\frac{\tau }{2}%
\sum_{\alpha }\mathbf{F}_{\alpha }^{\dag }\mathbf{F}_{\alpha }, \\
\mathbf{W}_{\beta >0} &=&\sqrt{\tau }\mathbf{F}_{\beta },
\end{eqnarray}%
and to the same order we also have the normalization condition
\begin{eqnarray}
\sum_{\beta =0}\mathbf{W}_{\beta }^{\dagger }\mathbf{W}_{\beta }=\mathbf{I}.
\end{eqnarray}%
Thus the Lindblad equation has been recast as a Kraus operator sum (\ref%
{Kraus}), but only to first order in $\tau $, the coarse-graining time scale
for which the Markovian approximation is valid \cite{Lidar:CP01}. This
implies a measurement interpretation, wherein the system state is $\rho
_{S}(t+\tau )=\mathbf{W}_{\beta }\rho (t)\mathbf{W}_{\beta }^{\dagger
}/p_{\beta }$ (to first-order in $\tau $) with probability $p_{\beta }=%
\mathrm{Tr}[\mathbf{W}_{\beta }\rho (t)\mathbf{W}_{\beta }^{\dagger }]$.
This happens because the bath functions as a probe coupled to the system
while being subjected to a quasi-continuous series of measurements at each
infinitesimal time interval $\tau $ \cite{ShabaniLidar:05}. The result is
the well-known quantum jump process \cite{Dalibard:92,Gisin:92,Plenio:98},
wherein the measurement operators are $\mathbf{W}_{0}\approx \exp (-i\tau
\mathbf{H}_{\mathrm{c}})$, the \textquotedblleft
conditional\textquotedblright\ evolution, generated by the non-Hermitian
\textquotedblleft Hamiltonian\textquotedblright\
\begin{eqnarray}
\mathbf{H}_{\mathrm{c}}\equiv \mathbf{H}_{S}-\frac{i}{2}\sum_{\alpha }%
\mathbf{F}_{\alpha }^{\dag }\mathbf{F}_{\alpha },  \label{eq:H_c}
\end{eqnarray}%
and $\sqrt{\tau }\mathbf{F}_{\beta }$ (the \textquotedblleft
jump\textquotedblright ). Note that $\mathbf{H}_{S}$ is here meant to
include all renormalization effects due to the system-bath interaction,
e.g., a possible Lamb shift (see, e.g., Ref.~\cite{Lidar:CP01}). By a simple algebraic
rearrangement one can rewrite the Lindblad equation in the following form:%
\begin{eqnarray}
\dot{\rho}_{S}=-i(\mathbf{H}_{\mathrm{c}}\rho _{S}-\rho _{S}\mathbf{H}_{%
\mathrm{c}}^{\dag })+\sum_{\alpha }\mathbf{F}_{\alpha }\rho _{S}\mathbf{F}%
_{\alpha }^{\dag },  \label{eq:jump}
\end{eqnarray}%
where according to the above interpretation the first term generates
non-unitary dynamics, while the second is responsible for the quantum jumps.

Now recall the Markovian DFS condition derived in Refs.~\cite%
{Lidar:PRL98,Zanardi:98}:\ the Lindblad operators should have trivial action
on DF-states, as in Eq.~(\ref{Fa}), i.e., $\mathbf{F}_{\alpha }|i\rangle
=c_{\alpha }|i\rangle $. Viewed from the perspective of the quantum-jump
picture of Markovian dynamics, this implies that the jump operators do not
alter a DF-state, i.e., the term $\sum_{\alpha }\mathbf{F}_{\alpha }\rho _{S}%
\mathbf{F}_{\alpha }^{\dag }$ in Eq.~(\ref{eq:jump}) transforms $\rho _{S}$
to $\sum_{\alpha }|c_{\alpha }|^{2}\rho _{S}$ and thus has trivial action.

Given Eq.~(\ref{Fa}), the Lindblad operators can be written in block-form as
follows [Eq.~(\ref{Markov-DFS})]:%
\begin{equation}
\mathbf{F}_{\alpha }=\left(
\begin{array}{cc}
c_{\alpha }\mathbf{I} & \mathbf{A}_{\alpha } \\
\mathbf{0} & \mathbf{B}_{\alpha }%
\end{array}%
\right) ,
\end{equation}%
with the blocks on the diagonal corresponding once again to operators
restricted to $\mathcal{H}_{\mathrm{DFS}}$ and $\mathcal{H}_{\mathrm{DFS}%
^{\bot }}$. Note the appearance of the off-diagonal block $\mathbf{A}%
_{\alpha }$ mixing $\mathcal{H}_{\mathrm{DFS}}$ and $\mathcal{H}_{\mathrm{DFS%
}^{\bot }}$; its presence is permitted since the DFS condition (\ref{Fa})
gives no information about matrix elements of the form $\langle i|\mathbf{F}%
_{\alpha }|j^{\bot }\rangle $, with $|i\rangle \in \mathcal{H}_{\mathrm{DFS}%
} $ and $|j^{\bot }\rangle \in \mathcal{H}_{\mathrm{DFS}^{\bot }}$.

As observed in \cite{Lidar:PRL98}, one should in addition require that $%
\mathbf{H}_{S}$ does not mix DF states with non-DF ones. It turns out that
this condition is compatible with the case that the DF state is imperfectly
initialized (Definition~\ref{def:imp}). In this case, as shown in Appendix~\ref%
{app:Markov}, the following theorem holds:

\begin{mytheorem}
\label{th:DFS-Markov-imperfect}Assume imperfect initialization. Then a subspace $\mathcal{H}_{\mathrm{DFS}}$ of the total Hilbert space $%
\mathcal{H}$ is decoherence-free with respect to Markovian dynamics iff the
Lindblad operators $\mathbf{F}_{\alpha }$ and the system Hamiltonian $%
\mathbf{H}_{S}$ assume the block-diagonal form
\begin{equation}
\mathbf{H}_{S}=\left(
\begin{array}{cc}
\mathbf{H}_{\mathrm{DFS}} & \mathbf{0} \\
\mathbf{0} & \mathbf{H}_{\mathrm{DFS}^{\bot }}%
\end{array}%
\right) ,\text{ \ }\mathbf{F}_{\alpha }=\left(
\begin{array}{cc}
c_{\alpha }\mathbf{I} & \mathbf{0} \\
\mathbf{0} & \mathbf{B}_{\alpha }%
\end{array}%
\right) ,
\end{equation}%
where $\mathbf{H}_{\mathrm{DFS}}$ and $\mathbf{H}_{\mathrm{DFS}^{\bot }}$
are Hermitian, $c_{\alpha }$ are scalars, and $\mathbf{B}_{\alpha }$ are
arbitrary operators on $\mathcal{H}_{\mathrm{DFS}^{\bot }}$.
\end{mytheorem}

But, as is clear from the quantum jumps picture, in particular Eqs.~(\ref%
{eq:H_c}),(\ref{eq:jump}), there also exists a non-Hermitian term, which
appears not to be addressed properly by merely restricting $\mathbf{H}_{S}$.
Indeed, this is the case if one demands that the system state is perfectly
initialized into the DFS (Definition~\ref{def:per}). As shown in Appendix~\ref{app:Markov},
the full condition on the Hamiltonian term then is:%
\begin{equation}
\langle i|(-i\mathbf{H}_{S}+\frac{1}{2}\sum_{\alpha }\mathbf{F}_{\alpha
}^{\dag }\mathbf{F}_{\alpha })|k^{\bot }\rangle =0,\quad \forall i,k^{\bot },
\label{new-DFS}
\end{equation}%
where $|i\rangle \in \mathcal{H}_{\mathrm{DFS}}$, $|k^{\bot }\rangle \in
\mathcal{H}_{\mathrm{DFS}^{\bot }}$. Applying the DFS conditions (\ref{Markov-DFS}),(\ref{new-DFS}), the Lindblad equation (\ref{Lindblad})
reduces to the Schr\"{o}dinger-like equation~(\ref{DFS-Lindblad}). Combining
these results, we have:

\begin{mytheorem}
\label{th:DFS-Markov-perfect}Assume perfect initialization. Then a
subspace $\mathcal{H}_{\mathrm{DFS}}$ of the total Hilbert space $\mathcal{H}
$ is decoherence-free with respect to Markovian dynamics iff the Lindblad
operators $\mathbf{F}_{\alpha }$ and Hamiltonian $\mathbf{H}_{S}$ satisfy
\begin{eqnarray}
\mathbf{F}_{\alpha } &=& \left(
\begin{array}{cc}
c_{\alpha }\mathbf{I} & \mathbf{A}_{\alpha } \\
\mathbf{0} & \mathbf{B}_{\alpha }%
\end{array}%
\right) \\
\mathcal{P}_{\mathrm{DFS}}\mathbf{H}_{S}\mathcal{P}_{\mathrm{DFS}}^{\dag}
&=& -\frac{i}{2}
\sum_{\alpha }c_{\alpha }^{\ast }\mathbf{A}_{\alpha }.
\label{DFS-Condition}
\end{eqnarray}
\end{mytheorem}

Note that $\mathbf{H}_{S}$ (which, again, includes the Lamb shift) must
satisfy a more stringent constraint than previously noted due to the extra
condition on its off-diagonal block. This has implications in examples of
practical interest, as we next illustrate.

\subsubsection{Example (significance of the new condition on the off-diagonal
blocks of $\mathbf{H}_{S}$)}

\label{example}

We present an example meant to demonstrate how the new constraint, Eq.~(\ref%
{new-DFS}) [or, equivalently, Eq.~(\ref{DFS-Condition})], may lead to a different prediction than the old constraint, that
matrix elements of the type $\langle j^{\bot }|\mathbf{H}_{S}|i\rangle $,
with $|i\rangle \in \mathcal{H}_{\mathrm{DFS}}$ and $|j^{\bot }\rangle \in
\mathcal{H}_{\mathrm{DFS}^{\bot }}$, should vanish.

Consider a system of three qubits interacting with a common bath. The system
is under influence of the bath via: 1) Spontaneous emission from the highest
level $|111\rangle $ to the lower levels, 2) Dephasing of the first and the
second qubits. For simplicity we set the system and bath Hamiltonians, $%
\mathbf{H}_{S}$ and $\mathbf{H}_{B}$, to zero. The total Hamiltonian then
contains only the system-bath interaction:

\begin{eqnarray}
\mathbf{H}_{I}&=&\lambda _{1}(\mathbf{\sigma }_{1}^{z}+\mathbf{\sigma }%
_{2}^{z})\otimes \mathbf{B}+\lambda _{2}[(\mathbf{\sigma }_{1}^{-}+\mathbf{%
    \sigma }_{2}^{-}+ \mathbf{\sigma }_{3}^{-})\otimes \mathbf{b}^{\dagger } \notag \\
  &&+ (\mathbf{\sigma }_{1}^{+}+\mathbf{\sigma }_{2}^{+}+\mathbf{\sigma }%
_{3}^{+})\otimes \mathbf{b}],
\end{eqnarray}%
where
\begin{equation}
\mathbf{\sigma }_{1}^{-}=|001\rangle \langle 111|,\text{ }\mathbf{\sigma }%
_{2}^{-}=|010\rangle \langle 111|,\text{ }\mathbf{\sigma }%
_{3}^{-}=|100\rangle \langle 111|,
\end{equation}%
and $\mathbf{b}$ is a bosonic annihilation operator.

The corresponding Lindblad equation may be derived, e.g., using the method
developed in Ref.~\cite{Lidar:CP01}. It may then be shown that

\begin{eqnarray}
\mathcal{L}[\rho _{S}]=\frac{1}{2}\sum_{i=1}^{2}[\mathbf{F}_{i},\rho _{S}%
\mathbf{F}_{i}^{\dagger }]+[\mathbf{F}_{i}\rho _{S},\mathbf{F}_{i}^{\dagger
}],
\end{eqnarray}%
where the Lindblad operators are

\begin{eqnarray}
\mathbf{F}_{1} &=&\sqrt{d_{1}}(u_{11}\mathbf{K}_{1}+u_{12}\mathbf{K}_{2}),
\notag \\
\mathbf{F}_{2} &=&\sqrt{d_{2}}(u_{21}\mathbf{K}_{1}+u_{22}\mathbf{K}_{2}).
\end{eqnarray}%
Here $\mathbf{K}_{1}=\mathbf{\sigma }_{1}^{z}+\mathbf{\sigma }_{2}^{z}$, $%
\mathbf{K}_{2}=\mathbf{\sigma }_{1}^{-}+\mathbf{\sigma }_{2}^{-}+\mathbf{%
\sigma }_{3}^{-}$, and $\{d_{1},d_{2}\}$ are the eigenvalues of the
Hermitian matrix $\mathbf{A}=[a_{ij}]$ of coefficients in the
pre-diagonalized Lindblad equation, with the diagonalizing matrix denoted $%
\mathbf{U}=[u_{ij}]$.

Now let us find the DFS conditions under the assumption of perfect
initialization. The previously-derived Eq.~(\ref{Fa})\ yields that $%
\{|000\rangle ,|001\rangle \}$ is a DFS, since $\mathbf{K}_{2}$ annihilates
these states, and they are both eigenstates of $\mathbf{K}_{1}$ with an
eigenvalue of $+2$:
\begin{eqnarray}
\mathbf{F}_{1}|000\rangle &=& 2\sqrt{d_{1}}u_{11}|000\rangle , \quad \mathbf{F}%
_{2}|000\rangle =2\sqrt{d_{2}}u_{21}|000\rangle \notag \\
\mathbf{F}_{1}|001\rangle &=& 2\sqrt{d_{1}}u_{11}|001\rangle , \quad \mathbf{F}%
_{2}|001\rangle =2\sqrt{d_{1}}u_{11}|001\rangle . \notag \\
\end{eqnarray}
However, the new condition~(\ref{new-DFS}) tightens the situation. Choosing
as representatives the states $|001\rangle \in \mathcal{H}_{\mathrm{DFS}}$
and $|111\rangle \in \mathcal{H}_{\mathrm{DFS}^{\bot }}$, we find from Eq.~(\ref{new-DFS}):
\begin{eqnarray}
\langle 001|\sum_{\alpha =1}^{2}\mathbf{F}_{\alpha }^{\dag }\mathbf{F}%
_{\alpha }|111\rangle &=& 2d_{1}u_{11}^{\ast }u_{12}+2d_{2}u_{21}^{\ast
}u_{22} \notag \\
&=& 0.
\end{eqnarray}
Since $u_{11}^{\ast }u_{12}+u_{21}^{\ast }u_{22}=0$ (from unitarity of $%
\mathbf{U}$), we see that the new condition imposes the extra symmetry
constraint $d_{1}=d_{2}$. This example illustrate the importance of the new
condition, Eq.~(\ref{new-DFS}).

\subsection{Noiseless Subsystems}

We now consider again the more general setting of subsystems, rather than
subspaces.

\subsubsection{Completely Positive Maps}

Suppose the system Hilbert space can be decomposed as $\mathcal{H}_{S}=
\mathcal{H}_{\mathrm{NS}}\otimes \mathcal{H}_{\mathrm{in}}\oplus \mathcal{H}
_{\mathrm{out}}$, where $\mathcal{H}_{\mathrm{NS}}$ is the factor in which
quantum information will be stored. The subspace $\mathcal{H}_{\mathrm{out}}$
may itself have a tensor product structure, i.e., additional factors similar
to $\mathcal{H}_{\mathrm{NS}}$ may be contained in it [as in
  Eq.~(\ref{dfss-decomposition})], but we shall not be interested in
those other
factors since the direct sum structure implies that different noiseless
factors cannot be used simultaneously in a coherent manner. As in the DF
subspace case considered above, we allow for the most general situation of a
system that is \emph{not} necessarily initially DF. To make this
notion precise, let us generalize the definitions of the projector
$\mathcal{P}_{\mathrm{DFS}}$ and projection operators
$\mathcal{P}_{\mathrm{d}},\mathcal{P}_{\mathrm{d}^\bot}$ given in the
DFS case, as follows:
\begin{equation}
\mathcal{P}_{\mathrm{NS-in}}=\left(
\begin{array}{cc}
\mathbf{I}_{\mathrm{NS}}\otimes \mathbf{I}_{\mathrm{in}} &
\mathbf{0}
\end{array}
\right) ,
\end{equation}

\begin{equation}
\mathcal{P}_{\mathrm{d}}=\left(
\begin{array}{cc}
\mathbf{I}_{\mathrm{NS}}\otimes \mathbf{I}_{\mathrm{in}} & \mathbf{0} \\
\mathbf{0} & \mathbf{0}%
\end{array}%
\right) , \quad
\mathcal{P}_{\mathrm{d}^\bot}=\left(
\begin{array}{cc}
\mathbf{0} & \mathbf{0} \\
\mathbf{0} & \mathbf{I}_{\mathrm{NS}}\otimes \mathbf{I}_{\mathrm{in}}
\end{array}%
\right)
\end{equation}
There is no risk of confusion in using the DFS notation,
$\mathcal{P}_{\mathrm{d}}$, for the NS case, as the DFS case is
obtained when $\mathbf{I}_{\mathrm{in}}$ is a scalar.

The system density matrix takes the corresponding block form
\begin{equation}
\rho _{S}=\left(
\begin{array}{cc}
\rho _{\mathrm{NS-in}} & \rho^\prime \\
\rho^{\prime \dag } & \rho _{\mathrm{out}}
\end{array}
\right) .
\label{eq:rho-NS-blocks}
\end{equation}

\begin{mydefinition}
  \label{def:NS}
Let the system Hilbert space $\mathcal{H}_S$ decompose as $\mathcal{H}_{S}=
\mathcal{H}_{\mathrm{NS}}\otimes \mathcal{H}_{\mathrm{in}}\oplus \mathcal{H}
_{\mathrm{out}}$, and partition the system state $\rho_{S}$ accordingly into
blocks, as in Eq.~(\ref{eq:rho-NS-blocks}). Assume $\rho_{\mathrm{NS-in}}(0)=
\mathcal{P}_{\mathrm{NS-in}}\rho _{S}(0)\mathcal{P}_{\mathrm{NS-in}}^{\dag }\neq
\mathbf{0}$. Then the factor $\mathcal{H}_{\mathrm{NS}}$ is called a decoherence-free (or
noiseless) subsystem if the following condition holds:
\begin{eqnarray}
  \mathrm{Tr}_{\mathrm{in}}\{\rho_{\mathrm{NS-in}}(t)\} =
  \mathbf{U}_{\mathrm{NS}}\mathrm{Tr}_{\mathrm{in}}\{\rho_{\mathrm{NS-in}}(0)\}\mathbf{U}_{\mathrm{NS}}^{\dagger},
\label{NS-Kraus-Def}
\end{eqnarray}
where $\mathbf{U}_{\mathrm{NS}}$ is a unitary matrix acting on $\mathcal{H}_{\mathrm{NS}}$.
\end{mydefinition}

\begin{mydefinition}
  \label{def:per-NS}
Perfect initialization (DF subsystems): $\rho^\prime =\mathbf{0}$
and $\rho_{\mathrm{out}} = \mathbf{0}$ in
Eq.~(\ref{eq:rho-NS-blocks}).
\end{mydefinition}

\begin{mydefinition}
  \label{def:imp-NS}
Imperfect initialization (DF subsystems): $\rho^\prime$ and/or $\rho_{\mathrm{out}}$ in Eq.~(\ref{eq:rho-NS-blocks}) are non-vanishing.
\end{mydefinition}

According to Definition~\ref{def:NS}, a quantum state encoded into the $\mathcal{H}_{%
\mathrm{NS}}$ factor at some time $t$ is unitarily related to the $t=0$
state. The factor $\mathcal{H}_{\mathrm{in}}$ is unimportant, and hence is
traced over. Clearly, a NS reduces to a DF subspace when $\mathcal{H}_{%
\mathrm{in}}$ is one-dimensional, i.e., when $\mathcal{H}_{\mathrm{in}}=\mathbb{C}$.

We now present the necessary and sufficient conditions for a NS and later we
show that the algebra-dependent definition, Eq.~(\ref{dfss-algebra}), is a
special case of this generalized form. In stating constraints on the form of
the Kraus operators, below, it is understood that in addition they must
satisfy the sum rule $\sum_{\alpha }\mathbf{E}_{\alpha }^{\dag }\mathbf{E}%
_{\alpha }=\mathbf{I}$, which we do not specify explicitly.

\begin{mytheorem}
\label{th:CP-NS-imperfect}Assume imperfect initialization. Then a subsystem $\mathcal{H}_{\mathrm{NS}}$ in the decomposition $%
\mathcal{H}_{S}=\mathcal{H}_{\mathrm{NS}}\otimes \mathcal{H}_{\mathrm{in}%
}\oplus \mathcal{H}_{\mathrm{out}}$ is decoherence-free (or noiseless) with
respect to CP maps iff the Kraus operators have the matrix representation%
\begin{eqnarray}
\mathbf{E}_{\alpha }=\left(
\begin{array}{cc}
\mathbf{U}\otimes \mathbf{C}_{\alpha } & \mathbf{0} \\
\mathbf{0} & \mathbf{B}_{\alpha }%
\end{array}%
\right)  \label{NS-Kraus}
\end{eqnarray}
\end{mytheorem}

\begin{mycorollary}
\label{NS}Assume perfect initialization. Then the Kraus operators
have the relaxed form%
\begin{eqnarray}
\mathbf{E}_{\alpha }=\left(
\begin{array}{cc}
\mathbf{U}\otimes \mathbf{C}_{\alpha } & \mathbf{A}_{\alpha } \\
\mathbf{0} & \mathbf{B}_{\alpha }%
\end{array}%
\right)  \label{NS-Kraus2}
\end{eqnarray}
\end{mycorollary}

We note that this result has been recently derived from an operator quantum error correction perspective in Ref.~\cite{Kribs:05}. Note again that there is a trade-off between the quality of
preparation and the amount of leakage that can be tolerated, a fact that was
not noted previously for subsystems, and has important experimental
implications.

As discussed above, the original definition of a NS was based on
representation theory of the error algebra. Here we have argued in favor of
a more comprehensive definition, based on the quantum channel picture. Let
us now state explicitly why our result is more general. Indeed, in the
algebraic approach one arrives at the representation (\ref{Kraus-DFSS}) of
the Kraus operators, namely $\mathbf{E}_{\alpha }=\bigoplus_{J\in \mathcal{J}%
}\mathbf{I}_{n_{J}}\otimes \mathbf{G}_{\alpha ,J}$. However, it is clear
from Eq.~(\ref{NS-Kraus2}) that our channel-based approach leads to a form
for the Kraus operators that includes this latter form as a special case,
since it allows for the off-diagonal block $\mathbf{A}_{\alpha }$. The
representation (\ref{Kraus-DFSS}) of the Kraus operators does agree with
Eq.~(\ref{NS-Kraus}), but in that case we do not need to assume
initialization inside the NS, so that again, our result is more general than
the algebraic one.

\subsubsection{Markovian Dynamics}

As in the CP-map based definition of a NS, we need to trace out the
$\mathcal{H}_{\mathrm{in}}$ factor, here in order to obtain the
dynamical equation for the subsystem factor:

\begin{eqnarray}
\frac{\partial \rho _{\mathrm{NS}}}{\partial t} &=&\frac{\partial \mathrm{Tr}
_{\mathrm{in}}\{\mathcal{P}_{\mathrm{NS-in}}\rho
  _{S}\mathcal{P}_{\mathrm{NS-in}}^{\dag }\}}{\partial t} \notag \\
&=& \mathrm{Tr}_{\mathrm{in}}\{\frac{\partial \mathcal{P
}_{\mathrm{NS-in}}\rho _{S}\mathcal{P}_{\mathrm{NS-in}}^{\dag }}{\partial t}\}
\notag \\
&=&\mathrm{Tr}_{\mathrm{in}}\{\mathcal{P}_{\mathrm{NS-in}}(-\frac{i}{\hbar }[
  \mathbf{H}_{S},\rho _{S}] +\frac{1}{2}\sum_{\alpha }2\mathbf{F}_{\alpha }\rho
_{S}\mathbf{F}_{\alpha }^{\dag } \notag \\
&& -\mathbf{F}_{\alpha }^{\dag }\mathbf{F}
_{\alpha }\rho _{S}-\rho _{S}\mathbf{F}_{\alpha }^{\dag }\mathbf{F}_{\alpha
})\mathcal{P}_{\mathrm{NS-in}}^{\dag }\}.
\label{markov-dfss}
\end{eqnarray}

\begin{mydefinition}
The factor $\mathcal{H}_{\mathrm{NS}}$ is called a decoherence-free (or
noiseless) subsystem under Markovian dynamics if a state subject to
Eq.~(\ref{markov-dfss}), undergoes continuous unitary evolution:
\begin{eqnarray}
\overset{\centerdot }{\rho }_{\mathrm{NS}}=i[\mathbf{M},\rho _{\mathrm{NS}}],
\label{eq:rhoNSdot}
\end{eqnarray}
where $\mathbf{M}$ is Hermitian.
\end{mydefinition}

Clearly, again, a NS reduces to a DF subspace when
$\mathcal{H}_{\mathrm{in}}$ is one-dimensional, i.e., when
$\mathcal{H}_{\mathrm{in}}=\mathbb{C}$.

Our goal is to find necessary and sufficient conditions such that
Eq.~(\ref{markov-dfss}) leads to Eq.~(\ref{eq:rhoNSdot}). In the case
of perfect initialization, since it does not involve
$\mathcal{H}_{\mathrm{out}}$, Eq.~(\ref{markov-dfss}) is meaningful
only if the system remains in the subspace $%
\mathcal{H}_{\mathrm{NS}}\otimes \mathcal{H}_{\mathrm{in}}$. An
analysis of Eq.~(\ref{markov-dfss}) reveals that this
leakage-prevention goal is achieved by imposing the constraints stated
in the following theorem, proven in Appendix~\ref{app:Markov}:

\begin{mytheorem}
\label{th:Markov-NS-perfect}Assume perfect initialization. Then
a subsystem $\mathcal{H}_{\mathrm{NS}}$ in the decomposition $\mathcal{H}%
_{S}=\mathcal{H}_{\mathrm{NS}}\otimes \mathcal{H}_{\mathrm{in}}\oplus
\mathcal{H}_{\mathrm{out}}$ is decoherence-free (or noiseless) with respect
to Markovian dynamics iff the Lindblad operators have the matrix
representation
\begin{eqnarray}
\mathbf{F}_{\alpha }\mathcal{=}\left(
\begin{array}{cc}
\mathbf{I}_{\mathrm{NS}}\otimes \mathbf{C}_{\alpha } & \mathbf{A}_{\alpha }
\\
\mathbf{0} & \mathbf{B}_{\alpha }%
\end{array}%
\right)  \label{dfss-markov3}
\end{eqnarray}%
and the system Hamiltonian (including a possible Lamb shift) has the matrix
representation%
\begin{eqnarray}
\mathbf{H}_{S}\mathcal{=}\left(
\begin{array}{cc}
\mathbf{H}_{\mathrm{NS}}\otimes \mathbf{I}_{\mathrm{in}}\mathbf{+I}_{\mathrm{%
NS}}\otimes \mathbf{H}_{\mathrm{in}} & \mathbf{H}_{2} \\
\mathbf{H}_{2}^{\dag } & \mathbf{H}_{3}%
\end{array}%
\right)  \label{dfss-markov4}
\end{eqnarray}%
where $\mathbf{H}_{\mathrm{in}}$ is constant along its diagonal, and where
\begin{eqnarray}
\mathbf{H}_{2}=-\frac{i}{2}\sum_{\alpha }\left( \mathbf{I}_{\mathrm{NS}%
}\otimes \mathbf{C}_{\alpha }^{\dag }\right) \mathbf{A}_{\alpha }.
\label{dfss-markov5}
\end{eqnarray}
\end{mytheorem}

Eqs.~(\ref{dfss-markov4}),(\ref{dfss-markov5}) are new additional
constraints on the Lindblad operators (compared to Ref.~\cite{Kempe:00})
which must be satisfied in order to find a NS.

If, on the other hand, we allow for imperfect initialization, we find a
different set of conditions:

\begin{mytheorem}
\label{th:NS-Markov-imperfect}Assume imperfect initialization. Then a subsystem $\mathcal{H}_{\mathrm{NS}}$ in the decomposition $%
\mathcal{H}_{S}=\mathcal{H}_{\mathrm{NS}}\otimes \mathcal{H}_{\mathrm{in}%
}\oplus \mathcal{H}_{\mathrm{out}}$ is decoherence-free (or noiseless) with
respect to Markovian dynamics iff the Lindblad operators have the matrix
representation%
\begin{eqnarray}
\mathbf{F}_{\alpha }=\left(
\begin{array}{cc}
\mathbf{I}_{\mathrm{NS}}\otimes \mathbf{C}_{\mathrm{in}}^{\alpha } & \mathbf{%
0} \\
\mathbf{0} & \mathbf{B}_{\alpha }%
\end{array}%
\right) ,
\end{eqnarray}%
and the system Hamiltonian (including a possible Lamb shift) has the matrix
representation%
\begin{eqnarray}
\mathbf{H}=\left(
\begin{array}{cc}
\mathbf{H}_{\mathrm{NS}}\otimes \mathbf{I}_{\mathrm{in}}\mathbf{+I}_{\mathrm{%
NS}}\otimes \mathbf{H}_{\mathrm{in}} & \mathbf{0} \\
\mathbf{0} & \mathbf{H}_{\mathrm{out}}%
\end{array}%
\right) .
\end{eqnarray}
\end{mytheorem}

\section{Performance of Quantum Algorithms over Imperfectly Initialized DFS\lowercase{s}}

\label{algo}

In this section we discuss applications of our generalized formulation of
DFSs to quantum algorithms. As mentioned above, a major obstacle to
exploiting decoherence-free methods is the unrealistic assumption of perfect
initialization inside a DFS. Removing this constraint enables us to perform
algorithms without perfect initialization, while not suffering from
information loss. We separate the role of an initialization error in the
algorithm (i.e., starting from an imperfect input state), from the effect of
noise in the output due to environment-induced decoherence. Thus we first
quantify an error entirely due to incorrect initialization ($\Delta _{%
\mathrm{leak}}$ below), then compare the DFS situations prior and post this
work, by relating them to $\Delta _{\mathrm{leak}}$.

1) Initialization error in the absence of decoherence: Assume no decoherence
at all, that the initial state is
\begin{eqnarray}
\rho ^{\mathrm{actual}}(0)\mathcal{=}\left(
\begin{array}{cc}
\rho _{1} & \rho _{2} \\
\rho _{2}^{\dag } & \rho _{3}%
\end{array}%
\right) ,  \label{eq:rho-ac}
\end{eqnarray}%
while the ideal input state is fully in the DFS:%
\begin{eqnarray}
\rho ^{\mathrm{ideal}}(0)\mathcal{=}\left(
\begin{array}{cc}
\rho & \mathbf{0} \\
\mathbf{0} & \mathbf{0}%
\end{array}%
\right) .  \label{eq:rho-id}
\end{eqnarray}%
Further assume that the algorithm is implemented via unitary
transformations $\mathbf{U=U}_{\mathrm{DFS}}\oplus \mathbf{I}_{\mathrm{DFS}^{\bot
}}$, applied to $\mathcal{H}_{\mathrm{DFS}}$. In general this will lead to
an output error in the algorithm, which can be quantified as
\begin{eqnarray}
\text{ }\Delta _{\mathrm{leak}} &\equiv &||\mathbf{U}\rho ^{\mathrm{actual}%
}(0)\mathbf{U}^{\dag }\mathbf{-U}\rho ^{\mathrm{ideal}}(0)\mathbf{U}^{\dag
}||  \notag \\
&=&\left\Vert \left(
\begin{array}{cc}
\mathbf{U}_{\mathrm{DFS}}(\rho _{1}-\rho )\mathbf{U}_{\mathrm{DFS}}^{\dag }
& \mathbf{U}_{\mathrm{DFS}}\rho _{2} \\
\rho _{2}^{\dag }\mathbf{U}_{\mathrm{DFS}}^{\dag } & \rho _{3}%
\end{array}%
\right) \right\Vert ,  \label{eq:Dleak}
\end{eqnarray}%
where $||\mathbf{\cdot }||$ denotes an appropriate operator norm. This error
appears not because of decoherence but because of an erroneous initial
state. This is a generic situation in quantum algorithms, which is not
special to the DFS case: Eq.~(\ref{eq:rho-ac}) is generic in the sense that
one can view the DFS block as the computational subspace, with the other
blocks representing additional levels (e.g., a qubit which is embedded in a
larger Hilbert space). Methods for correcting such deviations from the ideal
result exist (leakage elimination \cite{WuByrdLidar:02,ByrdLidarWuZanardi:05}), but are beyond the scope of this paper.

2) Initialization error in the presence of decoherence: Assume that the
input state is imperfectly initialized, as in Eq.~(\ref{eq:rho-ac}), and in
addition there is decoherence, i.e.,
\begin{equation}
\rho ^{\mathrm{actual}}(t)=\sum_{\alpha }\mathbf{E}_{\alpha }(t)\rho ^{%
\mathrm{actual}}(0)\mathbf{E}_{\alpha }^{\dagger }(t),
\end{equation}%
with the Kraus operators given by Eq.~(\ref{Mat-Rep}) [the form compatible
with decoherence-free evolution starting from $\rho ^{\mathrm{actual}}(0)$].
Prior to our work it was believed that for an imperfect initial state of the
form $\rho ^{\mathrm{actual}}(0)$, leakage due to the components $\rho _{2}$
and $\rho _{3}$ would cause non-unitary evolution of the DFS component.
Thus instead of an error $\mathbf{U}_{\mathrm{DFS}}(\rho _{1}-\rho )\mathbf{U%
}_{\mathrm{DFS}}^{\dag }$ in the DFS block of Eq.~(\ref{eq:Dleak}),
it was believed that one had $\mathcal{E}(\rho
_{1})-\mathbf{U}_{\mathrm{DFS}}\rho \mathbf{U}_{\mathrm{DFS}}^{\dag
}$ where $\mathcal{E}$ is an appropriate superoperator component.
This would have led to a reduced algorithmic fidelity, $\Delta
_{\mathrm{leak}}^{\prime }<\Delta _{\mathrm{leak}}$.
However, we now know that even for an initial state of the form $\rho ^{%
\mathrm{actual}}(0)$, when the Kraus operators are given by Eq.~(\ref{Mat-Rep})
the actual algorithmic fidelity is still given by $\Delta _{\mathrm{leak}}$,
since in fact the evolution of the DFS block is still unitary.

The above arguments apply when imperfect initialization is unavoidable
\emph{but one knows the component} $\rho _{1}$. A worse (though perhaps more
typical) scenario is one where not only is imperfect initialization
unavoidable, but one does not even know the component $\rho _{1}$. In this
case the above arguments apply in the context of algorithms that allow
arbitrary input states. Almost all the important examples of quantum
algorithms are now known to have a flexibility of this type:\ Grover's
algorithm \cite{Grover:96} was the first to be generalized to allow for
arbitrary input states, first pure \cite{BBBGL98,Lidar:PRA99Grover,Lidar:PRA01Grover}, then mixed \cite{Biham:02};
Shor's algorithm \cite{Shor:97} can run efficiently with a single pure qubit
and all other qubits in an arbitrary mixed state \cite{Parker:00}; a similar
result applies to a class of interesting physics problems, such as finding
the spectrum of a Hamiltonian \cite{Knill:98a}; the Deutsch-Josza \cite{Deutsch:92} algorithm was generalized to allow for arbitrary input states
\cite{Chi:01}, and a similar result holds for an algorithm that performs the
functional phase rotation (a generalized form of the conventional
conditional phase transform) \cite{Kim:02}. Most recently it was shown that
Simon's problem and the period-finding problem can be solved quantumly
without initializing the auxiliary qubits \cite{Chi:05}.

For algorithms that do not allow arbitrary input states, one could still
make use of the flexibility we have introduced into DFS state
initialization, provided it is possible to apply post-selection: one
modifies the output error of algorithm by observing whether the
measurement outcome came from the DFS block or not (this could be done,
e.g., via frequency-selective measurements, similar to the cycling
transition method used in trapped-ion quantum computing \cite{Wineland:98}).

\section{Decoherence Free Subspaces and Subsystems in non-Markovian Dynamics}

\label{ME}

\subsection{Decoherence Free Subspaces}

In Ref.~\cite{ShabaniLidar:05} a new class of non-Markovian master equations
was introduced. The following equation was derived as an analytically
solvable example of this class:

\begin{eqnarray}
\frac{\partial \rho _{S}}{\partial t}=-i[\mathbf{H}_{S},\rho _{S}]+\mathcal{L
}\int_{0}^{t}dt^{\prime }k(t^{\prime })\exp (\mathcal{L}t^{\prime })\rho
_{S}(t-t^{\prime }) \notag \\
\label{non-Markov}
\end{eqnarray}
where $\mathcal{L}$ is Lindblad super-operator and $k(t)$ represents the
memory effects of the bath. The Markovian limit is clearly recovered when $%
k(t)\propto \delta (t)$.\footnote{We note that Ref.~\cite{ShabaniLidar:05} contains a
small error: the Markovian limit is recovered for $k(t)=\delta (t)$ only if
the lower limit in Eq.~(\ref{non-Markov}) is $-t$. This change can easily be
applied to the derivation of Ref.~\cite{ShabaniLidar:05}.}

Some examples of physical systems which can be described by this master
equation are (i) a two-level atom coupled to a single cavity mode, wherein
the memory function is exponentially decaying, $k(t)=e^{-\lambda t}$ \cite%
{Breuer:book}, and (ii) a single qubit subject to telegraph noise in the
particular case that $||\mathcal{L}||\ll 1/t$, whence Eq.~(\ref{non-Markov})
reduces to $\dot{\rho}_{S}=\mathcal{L}\int_{0}^{t}dt^{\prime }k(t^{\prime
})\rho (t-t^{\prime })$ \cite{Daffer:03}. It is interesting to investigate
the conditions for a DFS in the case of dynamics governed by Eq.~(\ref{non-Markov}), and to compare the results with the Markovian limit, $k(t)\propto \delta (t)$. We defer proofs to Appendix~\ref{app:non-Markov}
and here present only the DFS-condition, stated in the following theorem (note that,
similarly to the Markovian case, we consider here a continuous-time DFS).

\begin{mytheorem}
\label{th:non-Markov-DFS-imperfect}Assume imperfect initialization. Then a subspace $\mathcal{H}_{\mathrm{%
DFS}}$ is decoherence free iff the system Hamiltonian $\mathbf{H}_{S}$ and
Lindblad operators $\mathbf{F}_{\alpha }$ have the matrix representation%
\begin{eqnarray}
\mathbf{H}_{S}=\left(
\begin{array}{cc}
\mathbf{H}_{\mathrm{DFS}} & \mathbf{0} \\
\mathbf{0} & \mathbf{H}_{\mathrm{DFS}^{\bot }}%
\end{array}%
\right) ,\text{ \ }\mathbf{F}_{\alpha }=\left(
\begin{array}{cc}
c_{\alpha }\mathbf{I} & \mathbf{0} \\
\mathbf{0} & \mathbf{B}_{\alpha }%
\end{array}%
\right)
\end{eqnarray}
\end{mytheorem}

These conditions are identical to those we found in the case of Markovian
dynamics with imperfect initialization -- cf. Theorem~\ref%
{th:DFS-Markov-imperfect}. This fact provides evidence for the robustness of
decoherence-free states against variations in the nature of the decoherence
process.

Interestingly, the conditions under the assumption of perfect initialization
differ somewhat when comparing the Markovian and non-Markovian cases:

\begin{mycorollary}
  \label{th:non-Markov-DFS-perfect}
Assume perfect initialization. Then a subspace
$\mathcal{H}_{\mathrm{DFS}} $ is decoherence free iff the system
Hamiltonian $\mathbf{H}_{S}$ and
Lindblad operators $\mathbf{F}_{\alpha }$ have the matrix representation%
\begin{eqnarray}
\mathbf{H}_{S} &=&\left(
\begin{array}{cc}
\mathbf{H}_{\mathrm{DFS}} & \mathbf{0} \\
\mathbf{0} & \mathbf{H}_{\mathrm{DFS}^{\bot }}%
\end{array}%
\right) , \\
\mathbf{F}_{\alpha } &=&\left(
\begin{array}{cc}
c_{\alpha }\mathbf{I} & \mathbf{A}_{\alpha } \\
\mathbf{0} & \mathbf{B}_{\alpha }%
\end{array}%
\right) \text{ and }\sum_{\alpha }c_{\alpha }^{\ast
}\mathbf{A}_{\alpha } = \mathbf{0}.
\end{eqnarray}
\end{mycorollary}

Compared to the Markovian case (Theorem~\ref{th:DFS-Markov-perfect}), the
difference is that now the off-diagonal blocks of the Hamiltonian must
vanish, whereas in the Markovian case we had the constraint
[Eq.~(\ref{DFS-Condition})]
$\mathcal{P}_{\mathrm{DFS}}\mathbf{H}_{S}\mathcal{P}_{\mathrm{DFS}}^{\dag}=-\frac{i}{2}
\sum_{\alpha }c_{\alpha }^{\ast }\mathbf{A}_{\alpha }$.

\subsection{Decoherence Free Subsystems}

We now consider the NS case. The dynamics governing a NS is derived by
tracing out $\mathcal{H}_{\mathrm{in}}$:
\begin{eqnarray}
\frac{\partial \rho _{\mathrm{NS}}}{\partial t} &=&\frac{\partial \mathrm{Tr}%
_{\mathrm{in}}\{\rho _{S}\}}{\partial t} = \mathrm{Tr}_{\mathrm{in}}\{\frac{%
\partial \rho _{S}}{\partial t}\} \notag \\
&=&\mathrm{Tr}_{\mathrm{in}}\{-i[\mathbf{H}_{S},\rho _{S}] \notag \\
&& +\mathcal{L}%
\int_{0}^{t}dt^{\prime }k(t^{\prime })\exp (\mathcal{L}t^{\prime })\rho
_{S}(t-t^{\prime })\}
\end{eqnarray}

\begin{mytheorem}
  \label{th:non-Markov-NS-imperfect}
Assume imperfect initialization.
Then a subsystem $\mathcal{H}_{\mathrm{NS}}$ in the decomposition $\mathcal{H%
}_{S}=\mathcal{H}_{\mathrm{NS}}\otimes \mathcal{H}_{\mathrm{in}}\oplus
\mathcal{H}_{\mathrm{out}}$ is decoherence-free (or noiseless) with respect
to non-Markovian dynamics [Eq.~(\ref{non-Markov})] iff the Lindblad
operators and the system
Hamiltonian have the matrix representation
\begin{eqnarray}
\mathbf{F}_{\alpha } &=&\left(
\begin{array}{cc}
\mathbf{I}_{\mathrm{NS}}\otimes \mathbf{C}_{\alpha } & \mathbf{0} \\
\mathbf{0} & \mathbf{B}_{\alpha }%
\end{array}%
\right) \\
\mathbf{H}_{S} &=&\left(
\begin{array}{cc}
\mathbf{H}_{\mathrm{NS}}\otimes \mathbf{I}_{\mathrm{in}}\mathbf{+I}_{\mathrm{%
NS}}\otimes \mathbf{H}_{\mathrm{in}} & \mathbf{0} \\
\mathbf{0} & \mathbf{H}_{\mathrm{out}}%
\end{array}%
\right) .
\end{eqnarray}
\end{mytheorem}

Note that this form is, once again, identical to the Markovian case with
imperfect initialization (cf. Theorem~\ref{th:NS-Markov-imperfect}).

However, as in the DFS case, the conditions are slightly different between
Markovian and non-Markovian dynamics if we demand perfect initialization:

\begin{mycorollary}
\label{th:non-Markov-NS-perfect}Assume perfect initialization. Then a
subsystem $\mathcal{H}_{\mathrm{NS}}$ in the decomposition $\mathcal{H}_{S}=%
\mathcal{H}_{\mathrm{NS}}\otimes \mathcal{H}_{\mathrm{in}}\oplus \mathcal{H}%
_{\mathrm{out}}$ is decoherence-free (or noiseless) with respect to
non-Markovian dynamics [Eq.~(\ref{non-Markov})] iff the Lindblad
operators and the system Hamiltonian have the matrix representation
\begin{eqnarray}
&&\mathbf{F}_{\alpha } =\left(
\begin{array}{cc}
\mathbf{I}_{\mathrm{NS}}\otimes \mathbf{C}_{\alpha } & \mathbf{A}_{\alpha }
\\
\mathbf{0} & \mathbf{B}_{\alpha }%
\end{array}
\right) , \\
&& \sum_{\alpha }(\mathbf{I}_{\mathrm{NS}}\otimes
\mathbf{C}_{\alpha}^{\dag })\mathbf{A}_{\alpha } = \mathbf{0}, \\
&& \mathbf{H} =\left(
\begin{array}{cc}
\mathbf{H}_{\mathrm{NS}}\otimes \mathbf{I}_{\mathrm{in}}\mathbf{+I}_{\mathrm{%
NS}}\otimes \mathbf{H}_{\mathrm{in}} & \mathbf{0} \\
\mathbf{0} & \mathbf{H}_{\mathrm{out}}%
\end{array}%
\right) .
\end{eqnarray}
\end{mycorollary}

\section{Summary and Conclusions}

We have revisited the concepts of decoherence-free subspaces and (noiseless)
subsystems (DFSs), and introduced definitions of DFSs that generalize
previous work. We have analyzed the conditions for the existence of DFSs in
the case of CP maps, Markovian dynamics, and (for the first time)
non-Markovian continuous-time dynamics. Our main finding implies
significantly relaxed demands on the preparation of decoherence-free states:
the initial state can be arbitrarily noisy. If, on the other hand, the
initial state is perfectly prepared, then almost arbitrary leakage
from outside the DFS into the DFS can be tolerated.

In the case of Markovian dynamics, if one demands perfect initialization,
our findings are of an opposite nature: we have shown that then an
additional constraint must be imposed on the system Hamiltonian, which
implies more stringent conditions for the possibility of manipulating a DFS
than previously believed. We have presented an example to illustrate this
fact.

We have also shown that the notion of noiseless subsystems, as originally
developed using an algebraic approach, admits a generalization when it is
instead developed from a quantum channel approach.

Our results have implications for experimental work on DFSs, and in
particular on quantum algorithms over DFSs \cite{Mohseni:02,Ollerenshaw:02}.
It is now known that a large class of quantum algorithms can tolerate almost
arbitrary preparation errors and still provide an advantage over their
classical counterparts
\cite{BBBGL98,Lidar:PRA99Grover,Lidar:PRA01Grover,Biham:02,Parker:00,Knill:98a,Chi:01,Kim:02,Chi:05}.
The relaxed preparation conditions for DFSs presented here are
naturally
compatible with this approach to quantum computation in noisy systems. This
should provide further impetus for the experimental exploration of quantum
computation over DFSs.

\begin{acknowledgments}
D.A.L. gratefully acknowledges financial support from Sloan
Foundation. This material is partially based on research sponsored
by the Defense Advanced Research Projects Agency under the QuIST
program and managed by the Air Force Research Laboratory (AFOSR),
under agreement F49620-01-1-0468.
\end{acknowledgments}

\appendix

\section{Proofs of Theorems and Corollaries}

Here we present proofs of all our results above. We shorten the calculations
by starting from the NS case and obtain the DFS conditions as a special case.

\subsection{CP Maps}

\label{app1}

\subsubsection{Arbitrary Initial State}

Assume the system evolution due to its interaction with a bath is described
by a CP map with Kraus operators $\{\mathbf{E}_{\alpha }\}$:
\begin{eqnarray}
\rho _{S}(t)=\sum_{\alpha }\mathbf{E}_{\alpha }\rho _{S}(0)\mathbf{E}%
_{\alpha }^{\dagger }.  \label{appendix1-A}
\end{eqnarray}%
Note that here $\rho _{S}$ is an operator on the entire system Hilbert space
$\mathcal{H}_{S}$, which we assume to be decomposable as $\mathcal{H}_{%
\mathrm{NS}}\otimes \mathcal{H}_{\mathrm{in}}\oplus \mathcal{H}_{\mathrm{out}%
}$. From the NS definition, Eq.~(\ref{NS-Kraus-Def}), we have
\begin{eqnarray}
\mathrm{Tr}_{\mathrm{in}}\{\mathbf{U}\otimes \mathbf{I}(\mathcal{P}_{\mathrm{NS-in}}\rho _{S}(0)\mathcal{P}_{\mathrm{NS-in}}^{\dag })\mathbf{U}^{\dagger }\otimes
  \mathbf{I}\}= \notag \\
\mathrm{Tr}_{\mathrm{in}}\{\sum_{\alpha }\left(
  \mathcal{P}_{\mathrm{NS-in}} \mathbf{E}_{\alpha }\right) \rho _{S}(0)( \mathbf{E}_{\alpha
}^{\dagger }\mathcal{P}_{\mathrm{NS-in}}^{\dag }) \}.  \label{appendix2-A}
\end{eqnarray}
Let us represent the Kraus operators in the same block-structure
matrix-form as that of the system state, i.e.,
corresponding to the decomposition $\mathcal{H}_{S}=\mathcal{H}%
_{\mathrm{NS}}\otimes \mathcal{H}_{\mathrm{in}}\oplus \mathcal{H}_{\mathrm{%
out}}$, where the blocks correspond to the subspaces $\mathcal{H}_{%
\mathrm{NS}}\otimes \mathcal{H}_{\mathrm{in}}$ (upper-left block) and $%
\mathcal{H}_{\mathrm{out}}$ (lower-right block). Then
\begin{eqnarray}
\rho _{S} &\mathcal{=}&\left(
\begin{array}{cc}
\rho _{1} & \rho _{2} \\
\rho _{2}^{\dag } & \rho _{3}%
\end{array}%
\right) ,  \label{eq:rho_gen} \\
\mathbf{E}_{\alpha } &=&\left(
\begin{array}{cc}
\mathbf{P}_{\alpha } & \mathbf{A}_{\alpha } \\
\mathbf{D}_{\alpha } & \mathbf{B}_{\alpha }%
\end{array}%
\right) ,  \label{eq:E_gen}
\end{eqnarray}%
with appropriate normalization constraints, considered below. Equation~(\ref{appendix2-A}) simplifies in this matrix form as
\begin{eqnarray}
\mathrm{Tr}_{\mathrm{in}}\{\mathbf{U}\otimes \mathbf{I}\rho _{1}\mathbf{U}%
  ^{\dagger }\otimes \mathbf{I}\} = \mathrm{Tr}_{\mathrm{in}}\{\sum_{\alpha }
\mathbf{P}_{\alpha }\rho _{1}\mathbf{P}_{\alpha }^{\dag } \notag \\
+\mathbf{P}_{\alpha
}\rho _{2}\mathbf{A}_{\alpha }^{\dag }+\mathbf{A}_{\alpha }\rho
_{2}^{\dagger }\mathbf{P}_{\alpha }^{\dag }+\mathbf{A}_{\alpha }\rho _{3}%
\mathbf{A}_{\alpha }^{\dag }\},
\label{appendix3-A}
\end{eqnarray}%
which must hold for arbitrary $\rho _{S}(0)$. To derive constraints on the
various terms we therefore consider special cases, which yield necessary
conditions. First, consider an initial state $\rho _{S}(0)$ such that $\rho
_{2}=\mathbf{0}$. Then, as the LHS of Eq.~(\ref{appendix3-A}) is independent
from $\rho _{3}$, the last term must vanish:%
\begin{eqnarray}
\sum_{\alpha }\mathbf{A}_{\alpha }\rho _{3}\mathbf{A}_{\alpha
}^{\dag }=\mathbf{0}\Longrightarrow \mathbf{A}_{\alpha }=\mathbf{0}.
\label{appendix4-A}
\end{eqnarray}%
Further assume $\rho _{1}=|i\rangle \langle i|\otimes |i^{\prime }\rangle
\langle i^{\prime }|$. Note that the partial matrix element $\langle
j^{\prime }|\mathbf{P}_{\alpha }|i^{\prime }\rangle $ is an operator on the $%
\mathcal{H}_{\mathrm{NS}}$ factor, $|i\rangle \langle i|$. Then Eq.~(\ref%
{appendix3-A}) reduces to%
\begin{eqnarray}
|i\rangle \langle i|=\sum_{\alpha ,j^{\prime }}\left[ \mathbf{U}^{\dagger
}\langle j^{\prime }|\mathbf{P}_{\alpha }|i^{\prime }\rangle \right]
|i\rangle \langle i|\left[ \langle i^{\prime }|\mathbf{P}_{\alpha }^{\dag
}|j^{\prime }\rangle \mathbf{U}\right] .
\end{eqnarray}%
Taking matrix elements with respect to $|i^{\bot }\rangle $, a state
orthogonal to $|i\rangle $, yields:%
\begin{eqnarray}
0 &=&\sum_{\alpha ,j^{\prime }}|\langle i^{\bot }|\left[ \mathbf{U}^{\dagger
}\langle j^{\prime }|\mathbf{P}_{\alpha }|i^{\prime }\rangle \right]
|i\rangle |^{2}  \notag \\
&\Longrightarrow &\langle i^{\bot }|\left[ \mathbf{U}^{\dagger
}\langle j^{\prime }|\mathbf{P}_{\alpha }|i^{\prime }\rangle \right]
|i\rangle =\mathbf{0},
\end{eqnarray}%
which, in turn implies that $\left[ \mathbf{U}^{\dagger }\langle j^{\prime }|%
\mathbf{P}_{\alpha }|i^{\prime }\rangle \right] |i\rangle $ is proportional
to $|i\rangle $, i.e.,
\begin{eqnarray}
\left[ \langle j^{\prime }|\mathbf{P}_{\alpha }|i^{\prime }\rangle \right]
|i\rangle \varpropto \mathbf{U}|i\rangle .  \label{appendix5-A}
\end{eqnarray}%
Since $|i^{\prime }\rangle ,|j^{\prime }\rangle $ are arbitrary this
condition implies that the submatrix $\mathbf{P}_{\alpha }$ must be of the
form $\mathbf{P}_{\alpha }=\mathbf{U}\otimes \mathbf{C}_{\alpha }$.
Substituting $\mathbf{P}_{\alpha }=\mathbf{U}\otimes \mathbf{C}_{\alpha }$
into Eq.~(\ref{appendix3-A}) we have $\mathrm{Tr}_{\mathrm{in}}\{\mathbf{U}%
\otimes \mathbf{I}\rho _{1}\mathbf{U}^{\dagger }\otimes \mathbf{I}\}=\mathrm{%
Tr}_{\mathrm{in}}\{\sum_{\alpha }\mathbf{U}\otimes \mathbf{C}_{\alpha }\rho
_{1}\mathbf{U^{\dag }}\otimes \mathbf{C}_{\alpha }^{\dag }\}$, so that
\begin{eqnarray}
\mathrm{Tr}_{\mathrm{in}}\{\rho _{1}\}=\mathrm{Tr}_{\mathrm{in}%
}\{\sum_{\alpha }\mathbf{I}_{\mathrm{NS}}\otimes \mathbf{C}_{\alpha }\rho
_{1}\mathbf{I}_{\mathrm{NS}}\otimes \mathbf{C}_{\alpha }^{\dag }\}.
\label{eq:TrinitiDF}
\end{eqnarray}%
Now suppose $\rho _{1}=\sum_{iji^{\prime }j^{\prime }}\lambda _{iji^{\prime
}j^{\prime }}|i\rangle \langle j|\otimes |i^{\prime }\rangle \langle
j^{\prime }|$; then from Eq.~(\ref{eq:TrinitiDF}) we find
\begin{eqnarray}
&&\sum_{iji^{\prime }}\lambda _{iji^{\prime }i^{\prime }}|i\rangle \langle
j| = \notag \\
&& \quad \sum_{iji^{\prime }j^{\prime }k^{\prime }\alpha }\lambda _{iji^{\prime
}j^{\prime }}|i\rangle \langle j|\text{ }\langle k^{\prime }|\mathbf{C}
_{\alpha }|i^{\prime }\rangle \langle j^{\prime }|\mathbf{C}_{\alpha }^{\dag
}|k^{\prime }\rangle .
\label{min-Norm}
\end{eqnarray}
Using $\sum_{k^{\prime }}|k^{\prime }\rangle \langle k^{\prime }|=\mathbf{I}%
_{\mathrm{in}}$, Eq.~(\ref{min-Norm}) becomes
\begin{eqnarray}
\sum_{iji^{\prime }}\lambda _{iji^{\prime }i^{\prime }}|i\rangle \langle
j|=\sum_{iji^{\prime }j^{\prime }}\lambda _{iji^{\prime }j^{\prime
}}|i\rangle \langle j|\langle j^{\prime }|\sum_{\alpha }\mathbf{C}_{\alpha
}^{\dag }\mathbf{C}_{\alpha }|i^{\prime }\rangle . \notag \\
\end{eqnarray}
It follows that
\begin{eqnarray}
\sum_{\alpha }\mathbf{C}_{\alpha }^{\dagger }\mathbf{C}_{\alpha }=\mathbf{I}
_{\mathrm{in}}.  \label{appendix6-A}
\end{eqnarray}

Next consider the normalization constraint $\sum_{\alpha }\mathbf{E}_{\alpha
}^{\dagger }\mathbf{E}_{\alpha }=\mathbf{I}$ for the Kraus operators,
together with the additional constraints we have derived ($\mathbf{A}%
_{\alpha }=\mathbf{0}$, $\mathbf{P}_{\alpha }=\mathbf{U}\otimes
\mathbf{C}_{\alpha }$):
\begin{eqnarray}
&& \sum_{\alpha }\mathbf{P}_{\alpha }^{\dagger }\mathbf{P}_{\alpha }+\mathbf{D}%
_{\alpha }^{\dagger }\mathbf{D}_{\alpha} = \mathbf{I}_{\mathrm{NS}} \otimes \mathbf{I}_{\mathrm{in}}
\notag \\
&& \Longrightarrow \mathbf{I}_{\mathrm{NS}}\otimes \sum_{\alpha }\mathbf{C}%
_{\alpha }^{\dagger }\mathbf{C}_{\alpha }+\sum_{\alpha }\mathbf{D}_{\alpha
}^{\dagger }\mathbf{D}_{\alpha }=\mathbf{I}_{\mathrm{NS}} \otimes
\mathbf{I}_{\mathrm{in}}. \notag \\
\end{eqnarray}
But, from Eq.~(\ref{appendix6-A}) we have $\sum_{\alpha }\mathbf{P}_{\alpha
}^{\dagger }\mathbf{P}_{\alpha }=\mathbf{I}_{\mathrm{NS}} \otimes \mathbf{I}_{\mathrm{in}}$. Therefore
$\mathbf{D}_{\alpha }=\mathbf{0}$.

Taking all these conditions together finalizes the matrix representation of
the Kraus operators as
\begin{eqnarray}
\mathbf{E}_{\alpha }=\left(
\begin{array}{cc}
\mathbf{U}\otimes \mathbf{C}_{\alpha } & \mathbf{0} \\
\mathbf{0} & \mathbf{B}_{\alpha }%
\end{array}%
\right) .  \label{appendix7-A}
\end{eqnarray}

For a scalar $\mathbf{C}_{\alpha }$ we recover the DFS condition (\ref{Mat-Rep}). These considerations establish the necessity of the
representation~(\ref{appendix7-A}); it is simple to show that this
representation is also sufficient, by substitution and checking that the NS
and DFS conditions are satisfied. Therefore we have proved Theorems~\ref{th:CP-DFS-imperfect} and \ref{th:CP-NS-imperfect}.

\subsubsection{Perfect Initialization}

We now prove Corollaries~\ref{DFS} and \ref{NS} for DF-initialized states of
the form $\rho _{S}(0)=\mathcal{P}_{\mathrm{d}}\rho _{S}(0)\mathcal{P}_{%
\mathrm{d}}$. Thus, we have to prove that $\mathbf{D}_{\alpha }=\mathbf{0}$
in Eq.~(\ref{eq:E_gen}).

When $\rho _{S}(0)=\mathcal{P}_{\mathrm{d}}\rho _{S}(0)\mathcal{P}_{\mathrm{d%
}}$ we have that $\rho _{2}=\mathbf{0}$ and $\rho _{3}=\mathbf{0}$ and Eq.~(\ref{appendix3-A}%
) reduces to
\begin{eqnarray}
\mathrm{Tr}_{\mathrm{in}}\{\mathbf{U}\otimes \mathbf{I}\rho _{1}\mathbf{U}%
^{\dagger }\otimes \mathbf{I}\}=\mathrm{Tr}_{\mathrm{in}}\{\sum_{\alpha }%
\mathbf{P}_{\alpha }\rho _{1}\mathbf{P}_{\alpha }^{\dag }\}.
\label{eq:initiDF}
\end{eqnarray}
The argument leading to the vanishing of the $\mathbf{A}_{\alpha }$
[Eq.~(\ref{appendix4-A})] then does not apply, and indeed the $\mathbf{A}%
_{\alpha }$ need not vanish. However, the arguments leading to $\mathbf{P}%
_{\alpha }=\mathbf{U}\otimes \mathbf{C}_{\alpha }$ and $\sum_{\alpha }\mathbf{P}
_{\alpha }^{\dagger }\mathbf{P}_{\alpha }=\mathbf{I}_{\mathrm{NS}}\otimes
\mathbf{I}_{\mathrm{in}}$ do apply. Hence $\mathbf{D}_{\alpha }=\mathbf{0}$.

\subsection{Markovian Dynamics}

\label{app:Markov}

\subsubsection{Arbitrary Initial State}

Consider Markovian dynamics

\begin{eqnarray}
\frac{\partial \rho _{S}}{\partial t}=-i[\mathbf{H}_{S},\rho
_{S}]+\sum_{\alpha }\mathbf{F}_{\alpha }\rho _{S}\mathbf{F}_{\alpha }^{\dag
} \notag \\
-\frac{1}{2}\mathbf{F}_{\alpha }^{\dag }\mathbf{F}_{\alpha }\rho _{S}-\frac{%
1}{2}\rho _{S}\mathbf{F}_{\alpha }^{\dag }\mathbf{F}_{\alpha },
\label{eq:MD}
\end{eqnarray}%
with the following matrix representation of the various operators:
\begin{eqnarray}
\rho _{S}&=&\left(
\begin{array}{cc}
\rho _{1} & \rho _{2} \\
\rho _{2}^{\dagger } & \rho _{3}%
\end{array}%
\right) ,\notag \\
\mathbf{H}_{S}&=& \left(
\begin{array}{cc}
\mathbf{H}_{1} & \mathbf{H}_{2} \\
\mathbf{H}_{2}^{\dag } & \mathbf{H}_{3}%
\end{array}%
\right) ,\,\,\mathbf{F}_{\alpha }=\left(
\begin{array}{cc}
\mathbf{P}_{\alpha } & \mathbf{A}_{\alpha } \\
\mathbf{D}_{\alpha } & \mathbf{B}_{\alpha }%
\end{array}%
\right) .  \notag \\
\label{eq:op-mat-rep}
\end{eqnarray}
Then we find the dynamics of the NS block to be
\begin{eqnarray}
&&\frac{\partial \rho _{\mathrm{NS}}}{\partial t} =\frac{\partial \mathrm{Tr}%
    _{\mathrm{in}}\{\rho _{1}\}}{\partial t} = \notag \\
&&-i\mathrm{Tr}_{\mathrm{in}}\{[%
\mathbf{H}_{1},\rho _{1}]\}-i\mathrm{Tr}_{\mathrm{in}}\{(\mathbf{H}_{2}\rho
_{2}^{\dagger }-\rho _{2}\mathbf{H}_{2}^{\dag })\} + \notag \\
&&\mathrm{Tr}_{\mathrm{in}}\{\sum_{\alpha }\mathbf{P}_{\alpha }\rho _{1}%
\mathbf{P}_{\alpha }^{\dag }+\mathbf{A}_{\alpha }\rho _{2}^{\dagger }\mathbf{%
P}_{\alpha }^{\dag }+\mathbf{P}_{\alpha }\rho _{2}\mathbf{A}_{\alpha }^{\dag
}+\mathbf{A}_{\alpha }\rho _{3}\mathbf{A}_{\alpha }^{\dag } \notag \\
&&-\frac{1}{2}\sum_{\alpha }(\mathbf{P}_{\alpha
}^{\dag }\mathbf{P}_{\alpha }+\mathbf{D}_{\alpha }^{\dag }\mathbf{D}_{\alpha
})\rho _{1}+(\mathbf{P}_{\alpha }^{\dag }\mathbf{A}_{\alpha }+\mathbf{D}%
_{\alpha }^{\dag }\mathbf{B}_{\alpha })\rho _{2}^{\dagger } \notag \\
&&-\frac{1}{2}\sum_{\alpha }\rho _{1}(\mathbf{P}%
_{\alpha }^{\dag }\mathbf{P}_{\alpha }+\mathbf{D}_{\alpha }^{\dag }\mathbf{D}%
_{\alpha })+\rho _{2}(\mathbf{A}_{\alpha }^{\dag }\mathbf{P}_{\alpha }+%
\mathbf{B}_{\alpha }^{\dag }\mathbf{D}_{\alpha })\}  \notag \\
\label{App-B1}
\end{eqnarray}%
The right-hand side of this equation must be independent of $\rho _{2}$ and $%
\rho _{3}$, for any matrices $\rho _{2}$ and $\rho _{3}$. Therefore the term
$\mathbf{A}_{\alpha }\rho _{3}\mathbf{A}_{\alpha }^{\dag }$ implies $\mathbf{%
A}_{\alpha }=\mathbf{0}$. Collecting the remaining terms acting on $\rho
_{2}^{\dagger }$ from the left yields $\mathrm{Tr}_{\mathrm{in}}\{(-i\mathbf{%
H}_{2}-\mathbf{D}_{\alpha }^{\dag }\mathbf{B}_{\alpha })\rho _{2}^{\dagger
}\}=\mathbf{0}$. Together we have
\begin{eqnarray}
\mathbf{A}_{\alpha }=\mathbf{0}, \quad i\mathbf{H}_{2}+\sum_{\alpha }\mathbf{D}
_{\alpha }^{\dag }\mathbf{B}_{\alpha }=\mathbf{0}.
\label{App-B2}
\end{eqnarray}%
This reduces Eq.~(\ref{App-B1}) to%
\begin{eqnarray}
&&\frac{\partial \rho _{\mathrm{NS}}}{\partial t} = \frac{\partial \mathrm{Tr}%
  _{\mathrm{in}}\{\rho _{1}\}}{\partial t} = \notag \\
&& -i\mathrm{Tr}_{\mathrm{in}}[\mathbf{%
H}_{1},\rho _{1}]+\mathrm{Tr}_{\mathrm{in}}\sum_{\alpha }\mathbf{P}_{\alpha
}\rho _{1}\mathbf{P}_{\alpha }^{\dag }  \notag \\
&&-\frac{1}{2}\mathrm{Tr}_{\mathrm{in}}\sum_{\alpha }\{(\mathbf{P}_{\alpha
}^{\dag }\mathbf{P}_{\alpha }+\mathbf{D}_{\alpha }^{\dag }\mathbf{D}_{\alpha
}),\rho _{1}\}  \label{App-B3}
\end{eqnarray}%
Consider the initial state $\rho _{1}=\rho _{\mathrm{NS}}\otimes |i^{\prime
}\rangle \langle i^{\prime }|$, with$\ |i^{\prime }\rangle \in \mathcal{H}_{%
\mathrm{in}}$:%
\begin{eqnarray}
&&\frac{\partial \rho _{\mathrm{NS}}}{\partial t} = -i[\langle i^{\prime }|%
    \mathbf{H}_{1}|i^{\prime }\rangle ,\rho _{\mathrm{NS}}] \notag \\
&&+\sum_{\alpha
}\langle j^{\prime }|\mathbf{P}_{\alpha }|i^{\prime }\rangle \rho _{\mathrm{%
NS}}\langle i^{\prime }|\mathbf{P}_{\alpha }^{\dag }|j^{\prime }\rangle \notag \\
&& -\frac{1}{2}\sum_{\alpha }\{\rho _{\mathrm{NS}},(\langle i^{\prime }|%
\mathbf{P}_{\alpha }^{\dag }|j^{\prime }\rangle \langle j^{\prime }|\mathbf{P%
}_{\alpha }|i^{\prime }\rangle \notag \\
&& +\langle i^{\prime }|\mathbf{D}_{\alpha
}^{\dag }|j^{\prime }\rangle \langle j^{\prime }|\mathbf{D}_{\alpha
}|i^{\prime }\rangle )\}  \label{App-B4}
\end{eqnarray}%
Let $\rho _{\mathrm{NS}}=|\psi \rangle \langle \psi |$\ with $\psi $
arbitrary and apply $\langle \psi ^{\bot }|...|\psi ^{\bot }\rangle $, such
that $\langle \psi ^{\bot }|\psi \rangle =0$, to Eq.~(\ref{App-B4}),
denoting $\mathbf{P}_{\alpha ,i^{\prime },j^{\prime }}\equiv \langle
j^{\prime }|\mathbf{P}_{\alpha }|i^{\prime }\rangle $:%
\begin{eqnarray}
\sum_{\alpha }|\langle \psi ^{\bot }|\mathbf{P}_{\alpha ,i^{\prime
},j^{\prime }}|\psi \rangle |^{2}=0.
\end{eqnarray}%
Since this identity must hold for all $\psi $ and $\psi ^{\bot }$, we find
that $\mathbf{P}_{\alpha ,i^{\prime },j^{\prime }}=c_{\alpha ,i^{\prime
},j^{\prime }}\mathbf{\mathbf{I}}_{\mathrm{NS}}$, which implies that $%
\mathbf{P}_{\alpha }=\mathbf{\mathbf{I}}_{\mathrm{NS}}\otimes \mathbf{%
\mathbf{C}}_{\mathrm{in}}^{\alpha }$. Moreover, by definition of a NS, there
exists a Hermitian matrix $\mathbf{H}_{\mathrm{NS}}$ such that $\rho _{%
\mathrm{NS}}$ obeys a Schr\"{o}dinger equation, $\partial \rho _{\mathrm{NS}%
}/\partial t=-i[\mathbf{H}_{\mathrm{NS}},\rho _{\mathrm{NS}}]$. Therefore
the non-Hermitian term $\sum_{\alpha }\mathbf{D}_{\alpha }^{\dag }\mathbf{D}%
_{\alpha }$ in Eq.~(\ref{App-B3}) must vanish, implying that $\mathbf{D}%
_{\alpha }=\mathbf{0}$.

Combining these results with Eq.~(\ref{App-B2}) yields
\begin{eqnarray}
\frac{\partial \mathrm{Tr}_{\mathrm{in}}\{\rho _{1}\}}{\partial t}&=&-i\mathrm{%
  Tr}_{\mathrm{in}}\{[\mathbf{H}_{1},\rho _{1}]\} \notag \\
&\equiv& -i[\mathbf{H}_{\mathrm{%
NS}},\rho _{\mathrm{NS}}]
\end{eqnarray}%
This identity can be realized iff $\mathbf{H}_{1}=\mathbf{H}_{\mathrm{NS}%
}\otimes \mathbf{I}_{\mathrm{in}}\mathbf{+I}_{\mathrm{NS}}\otimes \mathbf{H}%
_{\mathrm{in}}$. Therefore the NS conditions are obtained as%
\begin{eqnarray}
\mathbf{H} &=& \left(
\begin{array}{cc}
\mathbf{H}_{\mathrm{NS}}\otimes \mathbf{I}_{\mathrm{in}}\mathbf{+I}_{\mathrm{%
NS}}\otimes \mathbf{H}_{\mathrm{in}} & \mathbf{0} \\
\mathbf{0} & \mathbf{H}_{3}%
\end{array}%
\right) ,\notag \\
\mathbf{F}_{\alpha } &=& \left(
\begin{array}{cc}
\mathbf{\mathbf{I}}_{\mathrm{NS}}\otimes \mathbf{\mathbf{C}}_{\mathrm{in}%
}^{\alpha } & \mathbf{0} \\
\mathbf{0} & \mathbf{B}_{\alpha }%
\end{array}%
\right) .  \label{App-B5}
\end{eqnarray}%
The DFS condition is a special case of (\ref{App-B4}), with $\mathrm{\dim }(%
\mathcal{H}_{\mathrm{in}})=1$. This concludes the proof of Theorems~\ref%
{th:DFS-Markov-imperfect} and \ref{th:NS-Markov-imperfect}.

\subsubsection{Perfect Initialization}

Now consider perfect initialization:
\begin{eqnarray}
\rho _{S}=\left(
\begin{array}{cc}
\rho _{1} & =\mathbf{0} \\
=\mathbf{0} & =\mathbf{0}%
\end{array}%
\right) .
\end{eqnarray}%
This is just the case of an arbitrary initial state considered above, with $%
\rho _{2}=\mathbf{0}$ and $\rho _{3}=\mathbf{0}$ in
Eq.~(\ref{App-B1}). This then yields the dynamics of $\rho
_{\mathrm{NS}}$ as being given by Eq.~(\ref{App-B3}). Repeating the
derivation following Eq.~(\ref{App-B3}) we conclude again that
$\mathbf{D}_{\alpha }=\mathbf{0},$ $\mathbf{P}_{\alpha }\mathbf{=\mathbf{I}}_{\mathrm{%
NS}}\otimes \mathbf{\mathbf{C}}_{\mathrm{in}}^{\alpha }$ and $\mathbf{H}_{%
\mathrm{1}}=\mathbf{H}_{\mathrm{NS}}\otimes \mathbf{I}_{\mathrm{in}}\mathbf{%
+I}_{\mathrm{NS}}\otimes \mathbf{H}_{\mathrm{in}}$.

Note that Eq.~(\ref{App-B2}) now does not apply (it was obtained assuming
nonzero $\rho _{2},\rho _{3}$), i.e., we cannot conclude that $\mathbf{A}%
_{\alpha }$ and $\mathbf{H}_{2}$ vanish. This implies that that $\partial
\rho _{S}/\partial t$ has a non-zero off-diagonal elements, which, using the
master equation (\ref{eq:MD}), we calculate to be:
\begin{eqnarray*}
&\,&\text{upper right block:} \notag \\
&\,& \quad i\rho _{1}\mathbf{H}_{2}+\sum_{\alpha }
\mathbf{P}_{\alpha }\rho _{1}\mathbf{D}_{\alpha }^{\dag } -\frac{1}{2}\rho
_{1}(\mathbf{P}_{\alpha }^{\dag }\mathbf{A}_{\alpha }+\mathbf{D}_{\alpha
}^{\dag }\mathbf{B}_{\alpha }) \notag \\
&\,& \quad =i\rho _{1}\mathbf{H}_{2}-\frac{1}{2}\rho
_{1}\sum_{\alpha }(\mathbf{\mathbf{I}}_{\mathrm{NS}}\otimes \mathbf{\mathbf{C%
}}_{\mathrm{in}}^{\alpha \dag })\mathbf{A}_{\alpha } \\
&\,& \text{bottom right block: } \sum_{\alpha }\mathbf{D}_{\alpha }\rho
_{1}\mathbf{D}_{\alpha }^{\dag }=\mathbf{0}\text{.}
\end{eqnarray*}%
To prevent the appearance of corresponding off-diagonal blocks in $\rho _{S}$%
, we must therefore demand
\begin{eqnarray}
\mathbf{H}_{2}+\frac{i}{2}\sum_{\alpha }(\mathbf{\mathbf{I}}_{\mathrm{NS}%
}\otimes \mathbf{\mathbf{C}}_{\mathrm{in}}^{\alpha \dag
})\mathbf{A}_{\alpha }=\mathbf{0},
\end{eqnarray}%
which is Eq.~(\ref{dfss-markov5}). The DFS case is obtained with $\mathrm{%
\dim }(\mathcal{H}_{\mathrm{in}})=1$. This concludes the proof of Theorems~%
\ref{th:DFS-Markov-perfect} and \ref{th:Markov-NS-perfect}.

\subsection{Non-Markovian Dynamics}

\label{app:non-Markov}

The derivation of the conditions for decoherence-freeness in the case of
non-Markovian dynamics is somewhat different from the other two cases we
have considered, because of the appearance of the nonlocal-in-time integral
in the master equation:

\begin{eqnarray}
\frac{\partial \rho _{S}}{\partial t}=-i[\mathbf{H}_{S},\rho _{S}]+\mathcal{L%
}\int_{0}^{t}dt^{\prime }k(t^{\prime })\exp (\mathcal{L}t^{\prime })\rho
_{S}(t-t^{\prime }) \notag \\
\label{App-C1}
\end{eqnarray}%
In order to find necessary conditions on the structure of $\mathbf{H}_{S}$
and $\mathcal{L}$ consider the case of small $t$, expand%
\begin{eqnarray}
\rho _{S}(t)=\sum_{n=0}t^{n}\rho _{S}^{(n)}(0),\ \
k(t)=\sum_{m=0}t^{m}k^{(m)}(0),
\end{eqnarray}%
and substitute into Eq.~(\ref{App-C1}). The constant ($t^{0}$) term yields
\begin{eqnarray}
\rho _{S}^{(1)}(0)=-i[\mathbf{H}_{S},\rho _{S}(0)]\text{.}
\end{eqnarray}%
The terms involving $t^{1}$ yield, after Taylor-expanding $\exp (\mathcal{L}%
t^{\prime })$:
\begin{eqnarray}
2\rho _{S}^{(2)}(0)=-i[\mathbf{H}_{S},\rho _{S}^{(1)}(0)]+k(0)\mathcal{L}%
\rho _{S}(0)\text{.}
\end{eqnarray}%
Thus the solution of Eq.~(\ref{App-C1}) up to first and second order in time
is:%
\begin{eqnarray}
\rho _{S}(t) &=&\rho _{S}(0)-it[\mathbf{H}_{S},\rho _{S}(0)]+O(t^{2}),
\label{App-C2} \\
\rho _{S}(t) &=&\rho _{S}(0)-it[\mathbf{H}_{S},\rho _{S}(0)] \notag \\
&& -\frac{t^{2}}{2}
\{-[\mathbf{H}_{S},[\mathbf{H}_{S},\rho _{S}(0)]]+k(0)\mathcal{L}\rho
_{S}(0)\}+O(t^{3}).  \notag \\
\label{App-C3}
\end{eqnarray}

\subsubsection{Arbitrary Initial State}

Consider once again the matrix representations as in Eq.~(\ref{eq:op-mat-rep}%
). Substituting these expressions into the first order equation (\ref{App-C2}%
), the $\rho _{1}(t)$ block yields
\begin{eqnarray}
\rho _{\mathrm{NS}}(t) &=&\rho _{\mathrm{NS}}(0)-it\mathrm{Tr}_{\mathrm{in}%
}\{[\mathbf{H}_{\mathrm{1}},\rho _{1}(0)]\} \notag \\
&& -it\mathrm{Tr}_{\mathrm{in}}\{%
\mathbf{H}_{\mathrm{2}}\rho _{2}^{\dag }(0)-\rho _{2}(0)\mathbf{H}_{\mathrm{2%
}}^{\dag }\} \Longrightarrow  \notag \\
\mathbf{H}_{\mathrm{2}} &=& \mathbf{0},\text{ \ }\mathbf{H}_{\mathrm{1}%
}=\mathbf{H}_{\mathrm{NS}}\otimes \mathbf{I}_{\mathrm{in}}+\mathbf{I}_{%
  \mathrm{NS}}\otimes \mathbf{H}_{\mathrm{in}}.   \notag \\
\label{App-C4}
\end{eqnarray}
Continuing to second order, Eq.~(\ref{App-C3}), the NS block is found to be
\begin{eqnarray}
&\,&\rho _{\mathrm{NS}}(t) =\rho _{\mathrm{NS}}(0)-it[\mathbf{H}_{\mathrm{NS}%
    },\rho _{\mathrm{NS}}(0)] \notag \\
&\,&-\frac{t^{2}}{2}[\mathbf{H}_{\mathrm{NS}},[\mathbf{H%
}_{\mathrm{NS}},\rho _{\mathrm{NS}}(0)]] +\mathrm{Tr}_{\mathrm{in}}\{2k(0)\sum_{\alpha }\mathbf{P}_{\alpha }\rho
_{1}\mathbf{P}_{\alpha }^{\dag } \notag \\
&\,&+\mathbf{A}_{\alpha }\rho _{2}^{\dagger }%
\mathbf{P}_{\alpha }^{\dag }+\mathbf{P}_{\alpha }\rho _{2}\mathbf{A}_{\alpha
}^{\dag }+\mathbf{A}_{\alpha }\rho _{3}\mathbf{A}_{\alpha }^{\dag }  \notag
\\
&\,&-k(0)\sum_{\alpha }(\mathbf{P}_{\alpha }^{\dag }\mathbf{P}_{\alpha }+%
\mathbf{D}_{\alpha }^{\dag }\mathbf{D}_{\alpha })\rho _{1}+(\mathbf{P}%
_{\alpha }^{\dag }\mathbf{A}_{\alpha }+\mathbf{D}_{\alpha }^{\dag }\mathbf{B}%
_{\alpha })\rho _{2}^{\dagger }  \notag \\
&\,&-k(0)\sum_{\alpha }\rho _{1}(\mathbf{P}_{\alpha }^{\dag }\mathbf{P}%
_{\alpha }+\mathbf{D}_{\alpha }^{\dag }\mathbf{D}_{\alpha })+\rho _{2}(%
\mathbf{A}_{\alpha }^{\dag }\mathbf{P}_{\alpha }+\mathbf{B}_{\alpha }^{\dag }%
\mathbf{D}_{\alpha })\}. \notag \\
\end{eqnarray}%
The first three terms correspond to unitary evolution, but the remaining
terms are essentially identical to the case of Markovian dynamics and must
be made to vanish, just as in Eq.~(\ref{App-B1}). The same arguments used
there apply and consequently%
\begin{eqnarray}
\mathbf{F}_{\alpha }=\left(
\begin{array}{cc}
\mathbf{\mathbf{I}}_{\mathrm{NS}}\otimes \mathbf{\mathbf{C}}_{\mathrm{in}%
}^{\alpha } & \mathbf{0} \\
\mathbf{0} & \mathbf{B}_{\alpha }%
\end{array}%
\right) .  \label{App-C5}
\end{eqnarray}%
The conditions (\ref{App-C4}), (\ref{App-C5}) are necessary and sufficient
for unitary evolution of the NS block under our non-Markovian master
equation. The DFS case is obtained with $\mathrm{\dim }(\mathcal{H}_{\mathrm{%
in}})=1$. This concludes the proof of
Theorems~\ref{th:non-Markov-DFS-imperfect} and
\ref{th:non-Markov-NS-imperfect}. 

\subsubsection{Perfect Initialization}

Assume
\begin{eqnarray}
\rho _{S}(0)=\left(
\begin{array}{cc}
\rho (0) & \mathbf{0} \\
\mathbf{0} & \mathbf{0}%
\end{array}%
\right) ;
\end{eqnarray}%
then from the first order equation (\ref{App-C2}), the NS block is found to
satisfy%
\begin{eqnarray}
\rho _{\mathrm{NS}}(t) &=& \rho _{\mathrm{NS}}(0)-it\mathrm{Tr}_{\mathrm{in}}\{[%
  \mathbf{H}_{\mathrm{1}},\rho (0)]\} \Longrightarrow \notag \\
\mathbf{H}_{\mathrm{1}} &=& \mathbf{H}_{\mathrm{NS}}\otimes \mathbf{I}_{\mathrm{in}}+%
\mathbf{I}_{\mathrm{NS}}\otimes \mathbf{H}_{\mathrm{in}}.
\end{eqnarray}
To second order in time [Eq.~(\ref{App-C3})]:
\begin{eqnarray}
&\,& \rho _{\mathrm{NS}}(t) =\rho _{\mathrm{NS}}(0)-it[\mathbf{H}_{\mathrm{NS}%
},\rho _{\mathrm{NS}}(0)]  \notag \\
&\,& -\frac{t^{2}}{2}[\mathbf{H}_{\mathrm{NS}},[\mathbf{H%
}_{\mathrm{NS}},\rho _{\mathrm{NS}}(0)]] \notag \\
&\,& +\frac{t^{2}}{2}\mathrm{Tr}_{\mathrm{in}}\{-\mathbf{H}_{\mathrm{2}}\mathbf{%
H}_{\mathrm{2}}^{\dag }\rho (0)-\rho (0)\mathbf{H}_{\mathrm{2}}\mathbf{H}_{%
  \mathrm{2}}^{\dag } \notag \\
&\,& +2k(0)\sum_{\alpha }\mathbf{P}_{\alpha }\rho \mathbf{P}_{\alpha }^{\dag }-(%
\mathbf{P}_{\alpha }^{\dag }\mathbf{P}_{\alpha }+\mathbf{D}_{\alpha }^{\dag }%
\mathbf{D}_{\alpha })\rho (0) \notag \\
&\,& -\rho (0)(\mathbf{P}_{\alpha }^{\dag }\mathbf{P}%
_{\alpha }+\mathbf{D}_{\alpha }^{\dag }\mathbf{D}_{\alpha })\}, \notag \\
\end{eqnarray}
which is again similar to the Markovian case. Similar logic therefore yields
$\mathbf{H}_{2}=\mathbf{D}_{\alpha }=\mathbf{0}$, and hence%
\begin{eqnarray}
\mathbf{F}_{\alpha }=\left(
\begin{array}{cc}
\mathbf{I}_{\mathrm{NS}}\otimes \mathbf{C}_{\alpha } & \mathbf{A}_{\alpha }
\\
\mathbf{0} & \mathbf{B}_{\alpha }%
\end{array}%
\right) .
\end{eqnarray}%
Here we should notice that the density matrix $\rho _{S}(0)$ has an
off-diagonal element $\rho (0)\sum_{\alpha }(\mathbf{P}_{\alpha }^{\dag }%
\mathbf{A}_{\alpha }+\mathbf{D}_{\alpha }^{\dag }\mathbf{B}_{\alpha })=\rho
(0)\sum_{\alpha }\mathbf{P}_{\alpha }^{\dag }\mathbf{A}_{\alpha }$. This
term must vanish, for otherwise $\rho _{S}(t)$ has non-zero off-diagonal
elements. Summarizing, we have
\begin{eqnarray}
\mathbf{F}_{\alpha } &=&\left(
\begin{array}{cc}
\mathbf{I}_{\mathrm{NS}}\otimes \mathbf{C}_{\alpha } & \mathbf{A}_{\alpha }
\\
\mathbf{0} & \mathbf{B}_{\alpha }%
\end{array}%
\right) ,\quad \sum_{\alpha }(\mathbf{I}_{\mathrm{NS}}\otimes \mathbf{C%
}_{\alpha }^{\dag })\mathbf{A}_{\alpha }=0, \notag \\
\mathbf{H} &=&\left(
\begin{array}{cc}
\mathbf{H}_{\mathrm{NS}}\otimes \mathbf{I}_{\mathrm{in}}\mathbf{+I}_{\mathrm{%
NS}}\otimes \mathbf{H}_{\mathrm{in}} & \mathbf{0} \\
\mathbf{0} & \mathbf{H}_{\mathrm{out}}%
\end{array}%
\right) .
\end{eqnarray}%
The DFS case is obtained with $\mathrm{\dim }(\mathcal{H}_{\mathrm{in}})=1$.
This concludes the proof of Corollaries~\ref{th:non-Markov-DFS-perfect} and \ref{th:non-Markov-NS-perfect}.


\end{document}